\documentclass[12pt]{article}
\usepackage{graphicx}
\usepackage{pazha}
\tightenlines
\def\*{$^{*}$}

\def\nxrbloct{115}

\def\nxrbtot{1077}

\def\a{$^{\mbox{\small a}}$}
\def\b{$^{\mbox{\small b}}$}
\def\c{$^{\mbox{\small c}}$}
\def\d{$^{\mbox{\small d}}$}
\def\e{$^{\mbox{\small e}}$}

\def\ergs{erg~s$^{-1}$}

\begin{document}
\sloppypar
\begin{flushright}
{\it to be published in Astronomy Letters, Vol. 32, No. 7, pp. 456--477 (2006)}
\end{flushright}

\vspace{2cm}

\title
{\large\bf Hard X-ray Bursts Detected by the IBIS Telescope Onboard the INTEGRAL Observatory in 2003-2004} 

\author
{\bf I.V. Chelovekov\affilmark{1*}, S.A. Grebenev\affilmark{1*}, R.A. Sunyaev\affilmark{1,2}}

\affil{{\it $^1$ Space Research Institute, Russian Academy of
    Sciences, Profsoyuznaya 84/32, 117997 Moscow, Russia}\\ 
{\it $^2$ Max-Planck-Institut f\"ur Astrophysik, Karl-Schwarzschild-Str. 1, Postfach 1317, D-85741 Garching, Germany}\\ }

\vspace{3mm}

\received{16.12.2005}

\vspace{5mm}
\noindent
{\bf Abstract.}
All of the observations performed with the IBIS telescope onboard 
the INTEGRAL observatory during the first one and a half years of 
its in-orbit operation (from February 10, 2003, through July 2, 2004)
have been analyzed to find X-ray bursts. IBIS/ISGRI detector lightcurves 
total count rate in the energy range 15-25 keV revealed 1077 bursts 
of durations from $\sim5$ to $\sim500$ s detected with a high statistical 
significance (only one event over the entire period of observations 
could be detected by a chance with a probability of 20\%). Apart from the 
events associated with cosmic gamma-ray bursts (detected in the 
field of view or passed through the IBIS shield), solar flares, 
and activity of the soft gamma repeater SGR1806-20, we were able 
to localize 105 bursts and, with one exception, to identify them 
with previously known persistent or transient X-ray sources (96 were 
identified with known X-ray bursters). In one case, the burst 
source was a new burster in a low state. We named it IGR~J17364-2711. 
Basic parameters of the localized bursts and their identifications 
are presented in the catalog of bursts. Curiously enough, 61 bursts 
were detected from one X-ray burster -- \mbox{GX 354-0}. The statistical 
distributions of bursts in duration, maximum flux, and recurrence
time have been analyzed for this source. Some of the bursts observed 
with the IBIS/ISGRI telescope were also detected by the JEM-X telescope 
onboard the INTEGRAL observatory in the standard X-ray energy
range 3-20 keV.

{\bf DOI:\/} 10.1134/S1063773706070048 

\vfill
\noindent\rule{8cm}{1pt}\\
{$^*$ e-mails $<$chelovekov@hea.iki.rssi.ru$>$, $<$sergei@hea.iki.rssi.ru$>$}

\section*{ INTRODUCTION }
\noindent
The X-ray bursts detected by telescopes and detectors
onboard orbital astrophysical observatories
are mainly associated with thermonuclear explosions
on the surfaces of weakly magnetized accreting neutron
stars in low-mass X-ray binaries (type-I bursts).
Favorable conditions for the accumulation due to accretion of 
a fairly thick layer of matter on the stellar surface and the 
attainment of the pressure and temperature required for
thermonuclear ignition and explosive burning at its
base are created only in such objects. The luminosities
of such sources (bursters) at the burst time can
increase by two or three orders of magnitude relative
to their quiescent state, reaching a critical Eddington
level $L_{\rm c} \sim2.5\times10^{38} (M/M_{\sun}) 
(1-R_g/R)^{1/2} (1+X)^{-1}$ erg s$^{-1}$, where $M$ and $R$ 
are the mass and radius of the neutron star, $R_g=2GM/c^2$ 
is its gravitational radius, and $X$ is the hydrogen abundance.

Solar flares and cosmic gamma-ray bursts (GRBs),
events from sources of recurrent bursts (magnetars),
and individual events related to unsteady accretion in
binary systems (type-II bursts from low-mass X-ray
binaries and bursts from high-mass X-ray binaries
with accretion from an inhomogeneous stellar wind
of the companion star) are also observed in the Xray
energy range. Compared to all these events,
type-I X-ray bursts and their sources are of special,
independent interest to researchers, because their
observational properties are very peculiar and because
they carry direct information about the processes
near the surfaces of neutron stars under conditions
of superstrong gravitational field and pressure, ultrahigh
temperatures, and relativistic velocities. The
detection of type-I bursts, along with the detection
of coherent pulsations, serves as one of the most
important and most reliable criteria for identifying the
nature of the compact object (a neutron star) in X-ray
binaries.

The fact that bursts are commonly observed from
weak X-ray sources (or transients during their low
state) opens up a possibility for using them in searching
for hitherto unknown bursters with persistent X-ray 
fluxes below the level of reliable detection by currently
available wide-field X-ray instruments. Such
sources can be detected only during bursts, when
their X-ray luminosities increase by tens or hundreds
of times for a short time1. The INTEGRAL observatory
is equipped with unique wide-field telescopes
that allow sky fields with an area of more than 1000
square degrees to be simultaneously studied with
a flux sensitivity higher than 1 mCrab (over several
hours of observations) and an angular resolution
reaching several arcminutes. In addition, it devotes
up to 85\% of the physical time to continuous observations
of the region of the Galactic center and the
Galactic plane, where the bulk of the Galactic stellar
mass is concentrated. Therefore, INTEGRAL is best
suited for conducting such a search.

In this paper, to find X-ray bursts, we analyze
the time histories of the total count rate from the
ISGRI detector of the IBIS telescope onboard the
INTEGRAL orbital observatory in the energy range
15-25 keV based on observations during the first
one and a half years of its in-orbit operation (from
February 10, 2003, through July 2, 2004). For all of
the detected bursts, we attempted to localize (using
the IBIS sky mapping capabilities) and identify them
with persistent X-ray sources within the field of view.
We compiled a catalog of identified bursts and constructed
their time histories in a softer X-ray energy
range using data from the JEM-X monitor onboard
the INTEGRAL observatory if this was permitted by
the observational conditions. The maximum objective
of this study was an attempt to detect hitherto unknown
bursters or short-lived X-ray transients.

\section*{\bf INSTRUMENTS, OBSERVATIONS, AND DATA ANALYSIS}
\noindent

The INTEGRAL international orbital gamma-ray
observatory (Winkler et al. 2003) was placed in orbit
by a Russian PROTON launcher on October 17,
2002 (Eismont et al. 2003). There are four instruments
onboard the observatory: the SPI gammaray
spectrometer, the IBIS gamma-ray telescope, the
JEM-X X-ray monitor, and the OMC optical monitor.
Here, we use data from the upper detector layer
(ISGRI) of the IBIS telescope (Lebrun et al. 2003;
Ubertini et al. 2003) and the JEM-X monitor (Lund
et al. 2003).

The ISGRI detector layer of the IBIS telescope is
an array of $128 \times 128$ semiconductor CdTe elements
that are sensitive to photons in the energy range 15-200 
keV and that provide an energy resolution of\footnote{
The sources themselves can also be detected accidentally,
during deep observations or surveys carried out by very 
sensitive mirrorX-ray telescopes.However, their nature 
s much more difficult to identify in this case.}
$\sim$7\% at 100 keV. The IBIS telescope incorporates
a coded mask that allows this device to be used not only for
spectral and timing analyses of the emission, but also
for reconstructing the image of the sky region in the
$30\deg \times 30\deg$ field of view of the telescope (FWZR, the
fully coded area is $9\deg \times 9\deg$) with an angular resolution
of 12 arcmin (FWHM) and localizing X-ray and
gamma-ray sources to within 1-2 arcmin. The JEM-X
monitor is also a telescope with a coded aperture,
but it is adapted to the standard X-ray energy range
3-35 keV. A gas chamber that provides an energy
resolution of $\sim$16\% at 6 keV is used as the position
sensitive detector. The 13.2$\deg$ field of view (FWZR,
the fully coded area is 4.8$\deg$ in diameter) is limited by
a collimator. The angular resolution of 3.35 arcmin
allows bright sources to be localized more accurately
than those with the IBIS telescope.

Our analysis of the data from both telescopes was
based on the INTEGRAL standard data processing
software package (OSA), version 4.2. We searched
for bursts using the time profiles of the count rates for
all of the events recorded by the ISGRI detector in the
energy range 15-25 keV, irrespective of the photon
arrival direction. Based on the list of events compiled
at the GTI phase of the OSA 4.2 standard ISGRI
data processing procedure, we reproduced the time
histories of the count rates for each individual INTEGRAL
observation (corresponding to an individual
pointing) with a time resolution of 1, 5, and 10 s. A
total of 13777 individual INTEGRAL observations
from February 10, 2003, through July 2, 2004, were
analyzed. All these data are now publicly available in
the INTEGRAL archive. The duration of individual
observations reached $\sim$4--5 ks; the total exposure time
of all the observations used was 33.7 Ms. Figure 1
shows an exposure map for the IBIS observations
used here.

The derived lightcurves were analyzed for the presence 
of bursts. An excess of the signal-to-noise 
ratio $(S-\overline{S})/N$ above a preset
threshold $s_0$ in a particular time bin served as a
criterion for a burst. The attention was focused on
the count rate with a time step of 5 s that is most
sensitive (i.e., provides the maximum $S/N$ ratio in an
individual bin) to ordinary X-ray bursts from bursters.
Since the number of events recorded by the detector
in each time bin obeys a Poisson distribution, there
is a low, but finite probability $p(s_0)$ of recording a
random spike even in the absence of a real burst. To
filter out such random spikes, we set a fairly high
threshold, $s_0 = 5.1$. It ensures that the probability
of recording one random burst with $(S-\overline{S})/N\ga s_0$
in the entire time series being checked (with 
$M\sim 6.7\times 10^6$ time bins at a bin length of 5 s) 
does not exceed $p(s_0)\times M\simeq 20$\%. Since the total 
count rate of the detector depended on the emission from
all sources within the IBIS field of view, the mean
count rate $\overline{S}$ and the noise level 
$N=\overline{S^2}-\overline{S}^2$ in our formulas
were determined independently from each individual
pointing. The results of the described analysis were
also checked by a visual examination of the count rate
time histories. The same analysis was performed for
the time histories with a resolution of 10 s.

\begin{figure}[tp]
\centerline{\includegraphics[width=0.95\linewidth]{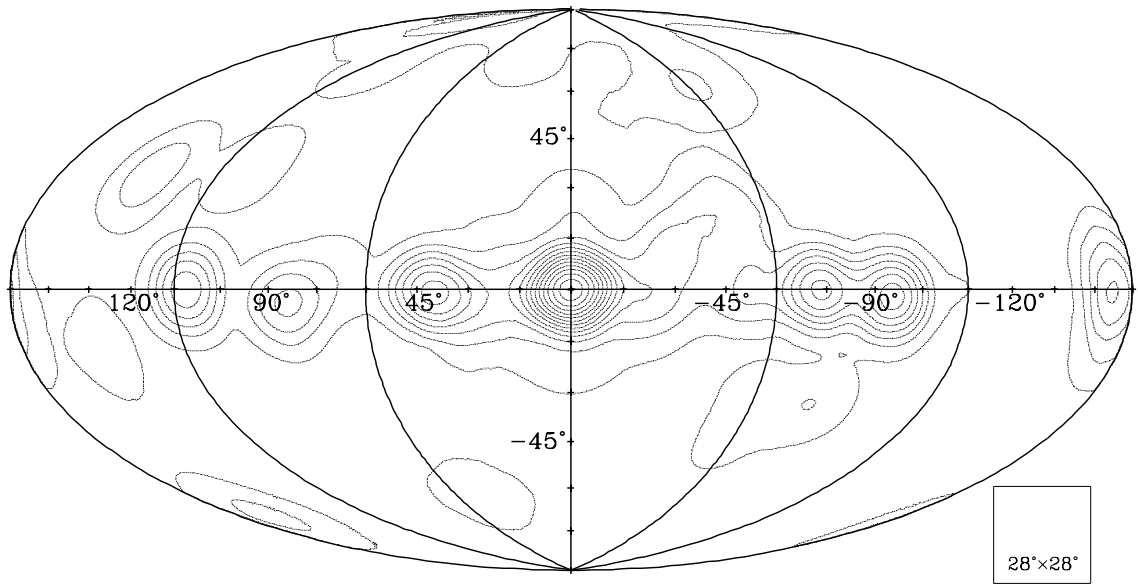}}
\caption{\rm Total exposure map of the IBIS pointings used to 
compile the catalog of bursts detected by the ISGRI detector in
2003-2004. The contours are given at 255 ks steps starting from 
the 73.4 ks exposure level. The exposure is seen to have
reached 3.39 Ms in the direction of the Galactic center. The IBIS 
field of view is schematically shown in the lower right corner.}
\label{fig:expo_map}
\end{figure}

For all of the detected bursts, we reconstructed
the images of the sky area in the IBIS/ISGRI field
of view (the IMA phase of the OSA 4.2 procedure)
accumulated with the same exposure at the burst
time and immediately before the burst (see Fig. 2).
We compared the statistical significance of detecting
sources in these images to reveal and identify the
burst source.

The entire set of procedures has much in common
with the IBAS system of the INTEGRAL observatory:
an automatic search for GRBs within the
IBIS/ISGRI field of view and a wide spread of notifications 
of them (Mereghetti et al. 2003). The differences
lie in the fact that the IBAS system (1) uses a
harder and wider energy ranges, (2) GRB-unrelated
events were ignored in the period 2003-2004, and (3)
the algorithms and programs were developed for realtime
automatic work with telemetry data.

\section*{\bf RESULTS}
\noindent
Our analysis of the IBIS/ISGRI detector lightcurves 
revealed 1077 bursts; the
sources of 115 of them were also detected in the
images of the sky region in the IBIS field of view.
Six such events were caused by cosmic GRBs that
occurred within the IBIS field of view. We were
able to associate other events, with one exception,
with known persistent X-ray sources. To clarify the
nature of 962 nonlocalized bursts, we performed their
cross correlation with the lists of GRBs and solar
flares recorded by the anticoincidence shield (ACS)
of the SPI gamma-ray spectrometer onboard the
INTEGRAL observatory and by the GOES satellites.
The results of this analysis are presented in Table 1.

\begin{table}[tb]
\caption{\rm Cross correlation of bursts detected by the 
IBIS/ISGRI telescope and events in the SPI/ACS and GOES 
experiments.\label{tab:ISGRI_GOES_ACS}} 

\begin{center}
\begin{tabular}{c|c|c|c|c|c} \hline

{Instrument}&
{ISGRI}&
{ISGRI\a}&
{ISGRI\b +}&
{ISGRI\b +}&
{ISGRI\b +}\\
{}&
{}&
{}&
{GOES}&
{ACS}&
{GOES+ACS}\\
\hline 
Bursts&\nxrbtot&\nxrbloct&145&86&17\\
\hline

\end{tabular} \\ [5mm]

\end{center}
\a Localized and identified ISGRI bursts.\\ 
\b Nonlocalized ISGRI bursts.\\ 
\end{table}

Table 2 lists the X-ray bursts detected by the
IBIS/ISGRI detector and identified with X-ray bursters.
Its columns give the burst source name, the maximum
flux time, the burst duration $T$, the maximum
flux determined from a time bin $\delta T$ of 1, 2, 3, or 5 s,
and the statistical significance of detecting the burst
source: (a) in the image obtained during the time bin
$\Delta T = 5$ or $10$ s near the burst maximum and (b) from
the time history of the ISGRI total count rate (from
the $S/N$ ratio in the same bin $\Delta T$). The values of $\delta T$
and $\Delta T$ used are given in columns 6 and 10 of the
table, respectively. In no case was the corresponding
persistent source detected in the image over the
period $\Delta T$ that preceded the burst onset by 1.5-2 
burst durations. All of the measurements were
performed in the energy range 15-25 keV. The table
also gives information about the success of detecting
a given burst by the JEM-X monitor. The profiles of
the bursts included in the table are shown in Fig. 3.

\begin{table}
\caption{\rm X-ray bursts detected by the IBIS/ISGRI telescope in 
the 15-25 keV energy range in 2003-2004.\label{tab:burst_cat}} 
\begin{center}
\vspace{-0.5cm}
\begin{tabular}{rlrcccrrrc@{}c}
\hline
\hline

{No.}&
{Source}&
\multicolumn{4}{c}{Burst maximum} &
\multicolumn{1}{c}{$T$\b,}&
\multicolumn{3}{c}{$(S-\overline{S})/N$}&
{JEM-X\e}\\
{}&
{}&
{Date,}&
{Time,}&
{Flux,}&
{$\delta T$\a,}&
{}&
{IM\c,}&
{LC\d,}&
{$\Delta T$,}&\\
{}&
{}&
{UTC}&
{h:m:s}&
{Crab}&
{s}&
\multicolumn{1}{c}{с}&
\multicolumn{1}{c}{$\sigma$}&
\multicolumn{1}{c}{$\sigma$}&
\multicolumn{1}{c}{с}&\\
\hline
\multicolumn{11}{c}{2003} \\
\hline
$ 1$&       GX~354-0&28.02&07:55:06&$2.52$&1&$ 13$&$12.2$&$12.9$& 5&+ \\
$ 2$&       GX~354-0&01.03&00:04:50&$2.48$&1&$ 13$&$12.6$&$ 9.4$& 5& \\
$ 3$&       GX~354-0&01.03&16:05:33&$3.04$&1&$ 11$&$10.9$&$ 7.8$& 5& \\
$ 4$&       GX~354-0&02.03&07:42:22&$3.04$&1&$  9$&$12.9$&$ 7.6$& 5& \\
$ 5$&    4U~1636-536&04.03&19:18:02&$1.13$&2&$ 10$&$ 7.9$&$ 6.6$& 5& \\
$ 6$&    4U~1702-429&09.03&21:51:13&$2.93$&1&$  9$&$12.1$&$14.5$& 5& \\
$ 7$&    4U~1608-522&09.03&22:35:05&$2.57$&1&$  8$&$10.8$&$ 9.0$& 5& \\
$ 8$&       GX~354-0&12.03&10:22:26&$2.70$&1&$  5$&$11.2$&$ 6.7$& 5& \\
$ 9$&    4U~1702-429&12.03&11:11:03&$3.09$&1&$  5$&$ 9.8$&$ 5.2$& 5& \\
$10$&    4U~1608-522&13.03&13:49:37&$2.61$&1&$  9$&$12.5$&$10.1$& 5& \\
$11$&    4U~1702-429&15.03&02:38:07&$2.44$&1&$  7$&$14.7$&$12.3$& 5&+\\
$12$&    4U~1702-429&15.03&18:22:49&$1.74$&2&$  5$&$ 7.8$&$ 5.8$& 5& \\
$13$&       GX~354-0&15.03&20:36:46&$1.22$&3&$  9$&$ 8.4$&$ 5.6$& 5& \\
$14$&       GX~354-0&03.04&08:40:18&$3.10$&1&$  6$&$11.6$&$ 8.2$& 5& \\
$15$&        Aql~X-1&06.04&07:42:15&$1.42$&2&$ 11$&$ 8.4$&$ 8.8$& 5&+\\
$16$&    4U~1724-307&06.04&18:32:31&$1.26$&2&$ 18$&$ 8.2$&$ 5.9$& 5& \\
$17$&       GX~354-0&06.04&19:45:29&$1.54$&2&$  8$&$ 7.2$&$ 6.5$& 5&+\\
$18$&       GX~354-0&07.04&03:26:31&$1.78$&1&$  6$&$ 8.1$&$ 7.5$& 5& \\
$19$&    4U~1636-536&11.04&18:13:18&$0.52$&6&$ 11$&$ 5.8$&$ 5.9$& 5& \\
$20$&    4U~1702-429&15.04&06:47:16&$2.54$&1&$  7$&$11.9$&$11.1$& 5& \\
$21$&     4U~1812-12&21.04&03:36:36&$2.55$&1&$ 14$&$11.8$&$ 5.6$&10& \\
$22$&     4U~1812-12&25.04&10:54:25&$3.60$&1&$  6$&$11.9$&$ 6.3$& 5& \\
$23$&    2S~0918-549&16.06&20:09:13&$3.61$&1&$ 25$&$12.8$&$ 5.5$& 5& \\
$24$&    4U~1702-429&18.08&10:05:10&$2.46$&1&$  7$&$11.9$&$ 9.5$& 5& \\
$25$&       GX~354-0&24.08&22:20:44&$1.69$&2&$  5$&$ 8.9$&$ 6.7$& 5& \\
$26$&SAX~J1712.6-3739&25.08&18:45:43&$1.78$&2&$ 9$&$10.1$&$ 5.1$& 5& \\
$27$&       GX~354-0&27.08&19:59:14&$1.82$&2&$  5$&$ 8.5$&$ 5.9$& 5& \\
$28$&       GX~354-0&28.08&01:24:04&$0.91$&5&$  7$&$ 7.1$&$ 8.8$& 5&+\\
$29$&       GX~354-0&28.08&06:01:30&$1.86$&2&$  7$&$ 7.8$&$ 5.7$& 5&+\\
$30$&       GX~354-0&29.08&14:31:29&$2.01$&2&$  5$&$10.0$&$ 6.4$& 5& \\
$31$&       GX~354-0&29.08&19:23:36&$2.19$&2&$  7$&$10.4$&$ 7.6$& 5& \\
$32$&       GX~354-0&31.08&15:54:18&$1.13$&3&$  7$&$ 7.6$&$10.1$& 5&+\\
$33$&       GX~354-0&03.09&03:26:34&$1.62$&2&$  7$&$10.6$&$ 9.5$& 5&+\\
\hline
\end{tabular}
\end{center}
$^a$ - The bin near the burst maximum used to calculate the flux.\\
$^b$ - The burst duration.\\
$^c$ - The source detection significance in the image during the bin $\Delta T$.\\
$^d$ - The source detection significance from the detector count rate in the bin $\Delta T$.\\
$^e$ - The bursts that were also detected by the JEM-X telescope.\\
\end{table}

\begin{table}
{Table 2:} Contd.
\begin{center}
\vspace{0.0cm}
\begin{tabular}{rlrcccrrrc@{}c}
\hline
\hline

{No.}&
{Source}&
\multicolumn{4}{c}{Burst maximum} &
\multicolumn{1}{c}{$T$\b,}&
\multicolumn{3}{c}{$(S-\overline{S})/N$}&
{JEM-X\e}\\
{}&
{}&
{Date,}&
{Time,}&
{Flux,}&
{$\delta T$\a,}&
{}&
{IM\c,}&
{LC\d,}&
{$\Delta T$,}&\\
{}&
{}&
{UTC}&
{h:m:s}&
{Crab}&
{s}&
\multicolumn{1}{c}{с}&
\multicolumn{1}{c}{$\sigma$}&
\multicolumn{1}{c}{$\sigma$}&
\multicolumn{1}{c}{с}&\\
\hline
\multicolumn{11}{c}{2003} \\
\hline
$34$&       GX~354-0&03.09&08:39:32&$1.49$&2&$  5$&$ 7.0$&$ 6.6$& 5&+\\
$35$&       GX~354-0&03.09&18:02:39&$1.08$&5&$  6$&$ 7.3$&$ 5.6$& 5& \\
$36$&     4U~1812-12&06.09&00:23:32&$4.04$&1&$ 19$&$11.8$&$ 9.3$& 5& \\
$37$&       GX~354-0&07.09&20:30:07&$1.55$&5&$  6$&$ 7.8$&$ 5.5$& 5& \\
$38$&       GX~354-0&08.09&13:41:36&$2.66$&1&$  7$&$ 8.8$&$ 7.1$& 5& \\
$39$&    4U~1724-307&08.09&18:48:30&$1.48$&1&$ 44$&$12.3$&$10.2$& 5&+\\
$40$&       GX~354-0&08.09&19:41:21&$2.63$&1&$  8$&$11.6$&$ 9.8$& 5&+\\
$41$&       GX~354-0&09.09&03:11:36&$2.95$&1&$  6$&$13.1$&$10.9$& 5& \\
$42$&       GX~354-0&09.09&16:28:54&$1.45$&3&$  7$&$ 7.7$&$ 8.6$& 5& \\
$43$&       GX~354-0&09.09&22:22:24&$2.58$&1&$  7$&$10.0$&$12.0$& 5&+\\
$44$&       GX~354-0&11.09&05:04:28&$3.92$&1&$  7$&$ 5.9$&$ 5.9$&10& \\
$45$&       GX~354-0&11.09&10:57:50&$1.99$&2&$  5$&$ 6.5$&$ 6.8$& 5& \\
$46$&       GX~354-0&11.09&21:59:51&$2.92$&1&$  5$&$12.3$&$13.6$& 5&+\\
$47$&       GX~354-0&12.09&03:12:25&$2.04$&1&$  4$&$ 8.5$&$ 9.3$& 5&+\\
$48$&       GX~354-0&12.09&09:22:40&$2.58$&1&$ 11$&$ 9.2$&$ 8.8$& 5&+\\
$49$&       GX~354-0&13.09&16:40:43&$1.03$&5&$  9$&$ 6.2$&$ 6.8$& 5& \\
$50$&       GX~354-0&13.09&22:28:39&$2.97$&1&$  6$&$ 9.2$&$ 6.8$& 5& \\
$51$&       GX~354-0&14.09&15:02:23&$2.72$&1&$  5$&$ 9.5$&$ 9.2$& 5&+\\
$52$&       GX~354-0&14.09&20:55:10&$2.17$&2&$  9$&$ 6.7$&$ 5.3$& 5& \\
$53$&       GX~354-0&15.09&09:40:18&$1.71$&3&$  6$&$ 6.1$&$ 7.4$& 5& \\
$54$&       GX~354-0&15.09&15:49:09&$3.10$&1&$  4$&$ 9.2$&$ 8.5$& 5& \\
$55$&   SLX~1735-269&15.09&17:42:29&$2.16$&1&$450$&$14.9$&$18.5$& 5&+\\
$56$&       GX~354-0&17.09&02:42:50&$2.79$&1&$  5$&$11.6$&$ 6.1$& 5& \\
$57$&       GX~354-0&17.09&08:58:31&$2.95$&1&$  7$&$ 9.8$&$ 7.8$& 5& \\
$58$&       GX~354-0&18.09&10:33:36&$2.42$&1&$  8$&$ 9.6$&$11.3$& 5& \\
$59$&       GX~354-0&19.09&16:12:10&$3.18$&1&$  5$&$ 9.1$&$14.5$& 5&+\\
$60$&       GX~354-0&20.09&05:40:37&$3.70$&1&$ 12$&$14.6$&$10.2$& 5& \\
$61$&       GX~354-0&20.09&23:47:03&$3.44$&1&$ 10$&$ 7.4$&$ 8.5$& 5& \\
$62$&       GX~354-0&21.09&14:07:40&$2.93$&1&$  8$&$11.8$&$15.4$& 5& \\
$63$&       GX~354-0&22.09&17:38:26&$2.81$&2&$  6$&$ 9.7$&$ 7.5$& 5& \\
$64$&       GX~354-0&23.09&02:16:11&$2.40$&1&$  7$&$14.6$&$11.2$& 5& \\
$65$&   SLX~1735-269&23.09&05:11:43&$1.04$&5&$  9$&$ 8.1$&$ 7.4$& 5& \\
$66$&       GX~354-0&23.09&10:53:38&$2.37$&1&$  5$&$ 7.7$&$ 7.5$& 5& \\
\hline
\end{tabular} 
\end{center}

$^a$ - The bin near the burst maximum used to calculate the flux.\\
$^b$ - The burst duration.\\
$^c$ - The source detection significance in the image during the bin $\Delta T$.\\
$^d$ - The source detection significance from the detector count rate in the bin $\Delta T$.\\
$^e$ - The bursts that were also detected by the JEM-X telescope.\\

\end{table}

\begin{table}
{Table 2:} Contd.
\begin{center}
\vspace{-0.5cm}
\begin{tabular}{rlrcccrrrc@{}c}
\hline
\hline

{No.}&
{Source}&
\multicolumn{4}{c}{Burst maximum} &
\multicolumn{1}{c}{$T$\b,}&
\multicolumn{3}{c}{$(S-\overline{S})/N$}&
{JEM-X\e}\\
{}&
{}&
{Date,}&
{Time,}&
{Flux,}&
{$\delta T$\a,}&
{}&
{IM\c,}&
{LC\d,}&
{$\Delta T$,}&\\
{}&
{}&
{UTC}&
{h:m:s}&
{Crab}&
{s}&
\multicolumn{1}{c}{с}&
\multicolumn{1}{c}{$\sigma$}&
\multicolumn{1}{c}{$\sigma$}&
\multicolumn{1}{c}{с}&\\
\hline
\multicolumn{11}{c}{2003} \\
\hline
$67$&       GX~354-0&23.09&18:15:09 &$2.61$&1&$  6$&$13.4$&$ 9.5$& 5& \\
$68$&   SLX~1735-269&23.09&23:13:11 &$1.05$&5&$  8$&$ 6.7$&$ 6.6$& 5& \\
$69$&       GX~354-0&24.09&03:52:12 &$3.18$&1&$ 10$&$13.0$&$ 8.9$& 5& \\
$70$&       GX~354-0&24.09&11:01:26 &$3.00$&1&$ 11$&$15.2$&$13.2$& 5&+\\
$71$&SAX~J1712.6-3739&24.09&14:00:09&$2.18$&2&$ 18$&$11.9$&$ 5.0$&10& \\
$72$&       GX~354-0&24.09&18:20:21 &$2.63$&1&$  5$&$ 8.0$&$ 5.9$& 5& \\
$73$&    4U~1608-522&26.09&02:38:55 &$3.73$&1&$ 14$&$16.2$&$12.1$& 5& \\
$74$&    4U~1608-522&26.09&15:34:51 &$3.79$&1&$ 11$&$21.4$&$18.9$& 5&+\\
$75$&    4U~1608-522&27.09&04:03:44 &$4.95$&1&$ 12$&$21.1$&$33.2$& 5&+\\
$76$&    4U~1812-12 &27.09&16:08:45 &$3.06$&1&$ 17$&$12.1$&$ 8.7$& 5& \\
$77$&       GX~354-0&04.10&22:06:42 &$1.68$&2&$ 12$&$ 6.9$&$ 6.8$& 5& \\
$78$&       GX~354-0&05.10&09:34:42 &$2.72$&1&$  6$&$ 9.1$&$ 5.0$& 5& \\
\hline
\multicolumn{11}{c}{2004} \\
\hline

$79$&       GX~354-0&17.02&04:47:50&$3.27$&1&$  7$&$14.4$&$11.0$& 5& \\
$80$&IGR~J17364-2711&17.02&14:41:30&$1.59$&1&$ 13$&$ 8.7$&$ 6.8$&10& \\
$81$&       GX~354-0&27.02&10:55:16&$3.08$&1&$  8$&$10.6$&$ 7.8$& 5&+\\
$82$&       GX~354-0&27.02&13:32:36&$1.53$&2&$  7$&$ 7.5$&$ 7.1$& 5&+\\
$83$&       GX~354-0&27.02&15:32:03&$2.11$&1&$  7$&$ 9.3$&$ 6.6$& 5&+\\
$84$&       GX~354-0&02.03&07:34:38&$1.52$&2&$  6$&$ 7.8$&$ 5.8$& 5& \\
$85$&         GX~3+1&02.03&09:25:34&$0.98$&2&$  7$&$ 7.5$&$ 6.4$& 5&+\\
$86$&    4U~1724-307&03.03&04:14:60&$1.40$&3&$  8$&$ 9.0$&$ 8.3$& 5&+\\
$87$&    4U~1608-522&20.03&20:59:32&$3.72$&1&$ 13$&$11.2$&$18.1$& 5& \\
$88$&    4U~1608-522&21.03&01:03:47&$1.83$&1&$ 10$&$10.4$&$ 9.3$& 5&+\\
$89$&        Aql~X-1&24.03&17:03:35&$1.58$&2&$  9$&$ 7.0$&$ 7.0$& 5& \\
$90$&       GX~354-0&29.03&02:40:47&$0.99$&3&$  5$&$ 6.5$&$ 5.6$& 5& \\
$91$&       GX~354-0&30.03&03:25:33&$1.25$&5&$  4$&$ 7.2$&$ 7.8$& 5& \\
$92$&   SLX~1744-299&30.03&03:37:46&$0.81$&5&$ 22$&$ 7.6$&$ 6.6$& 5& \\
$93$&    KS~1741-293&30.03&03:43:45&$0.88$&5&$  7$&$ 6.9$&$ 5.7$& 5& \\
$94$&       GX~354-0&31.03&03:09:05&$0.85$&5&$  4$&$ 6.4$&$ 5.5$& 5& \\
$95$&       GX~354-0&01.04&23:36:53&$0.99$&5&$  5$&$ 6.3$&$ 5.5$& 5& \\
$96$&        Aql~X-1&28.04&07:54:48&$1.96$&2&$  7$&$ 8.9$&$ 8.1$& 5& \\
$97$&        Aql~X-1&01.05&22:56:47&$1.89$&1&$ 10$&$ 7.3$&$ 8.6$& 5&+\\
\hline
\end{tabular}
\end{center}

$^a$ The bin near the burst maximum used to calculate the flux.\\
$^b$ The burst duration.\\
$^c$ The source detection significance in the image during the bin $\Delta T$.\\
$^d$ The source detection significance from the detector count rate in the bin $\Delta T$.\\
$^e$ The bursts that were also detected by the JEM-X telescope.\\
\end{table}

There are several multipeaked bursts among the
events listed in Table 2. For these bursts, column 4
gives the time of the maximum flux over the entire
burst, not in the first peak. To determine the burst
duration $T$, we analyzed the time history of the ISGRI
count rate in the corresponding session with
a resolution of 1~s. The burst onset and end were
defined as the times the count rate exceeded the
session-averaged level by 10\%. The maximum burst
flux was determined when analyzing the image of the
sky area in the IBIS field of view reconstructed with
the OSA 4.2 software package during $\delta T$. The fluxes
in Table 2 are in units of the IBIS/ISGRI flux from
the Crab Nebula, which allows them to be used for
physical estimations, for example, for estimating the
burst fluence.

\begin{figure}[hp]
\centerline{\includegraphics[width=0.7\linewidth]{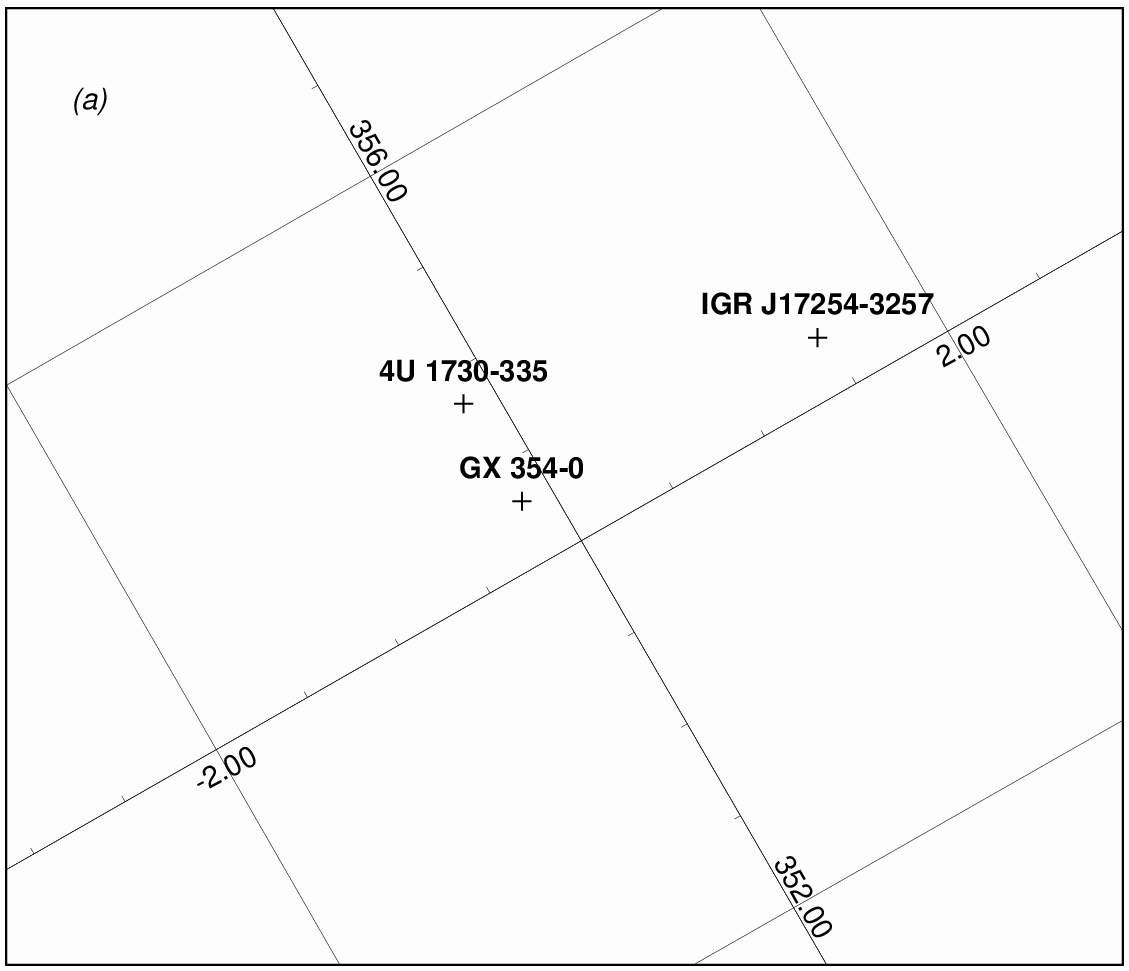}}
\centerline{\includegraphics[width=0.7\linewidth]{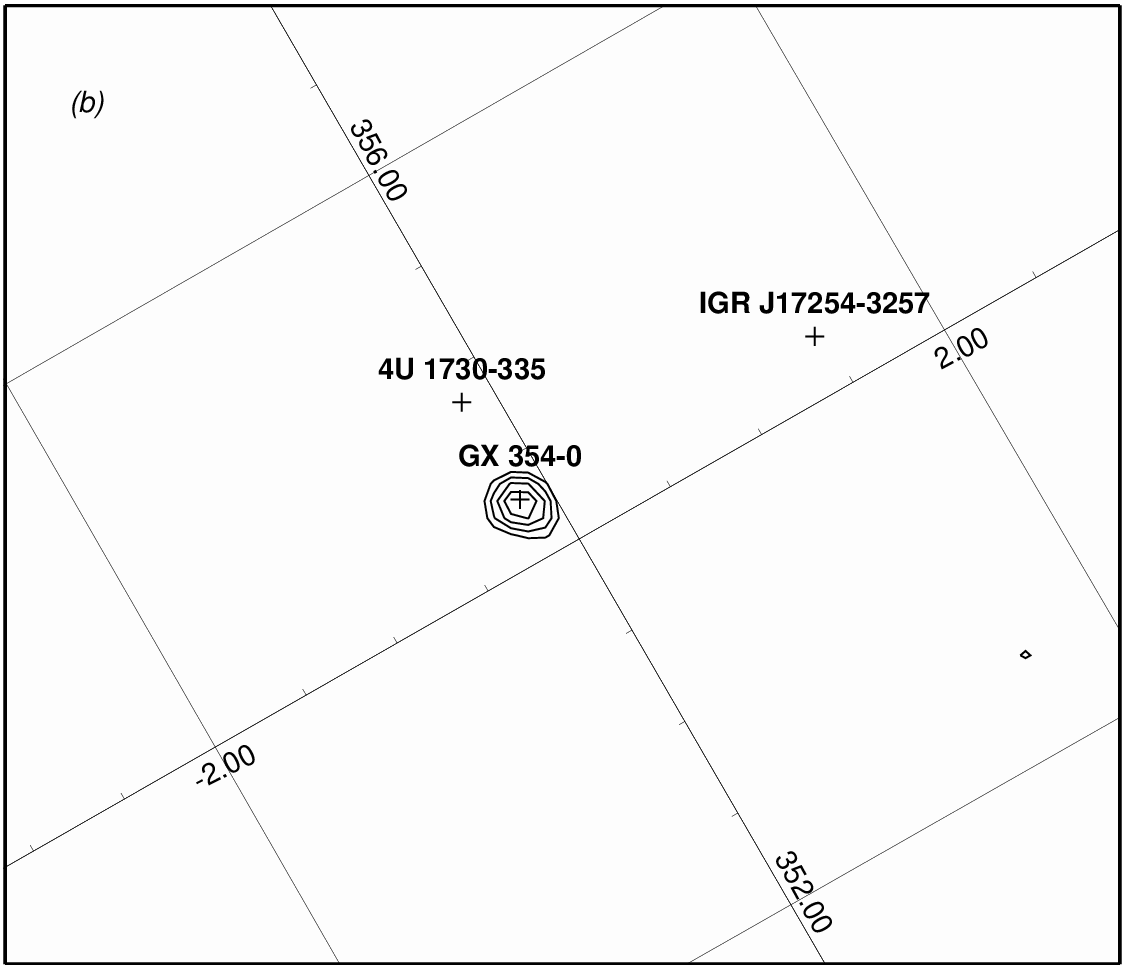}}
\caption{\rm Image of the sky region in the IBIS/ISGRI field
of view 10 s before the onset of the burst detected on
September 20, 2003, (a) and at its maximum (b). In
both cases, the exposure is 5 s and the energy range is
15--25 keV. The contours indicate the region of reliable
detection of the source at the signal-to-noise ratio $S/N =
4, 6.5, 9.1, 11.7$, and $14.3$ standard deviations (given with
a logarithmic step of 1.22).}
\label{fig:skyimage}
\end{figure}

\begin{figure}[hp]
\vspace{-5.5cm}
\centerline{\includegraphics[width=1.5\linewidth]{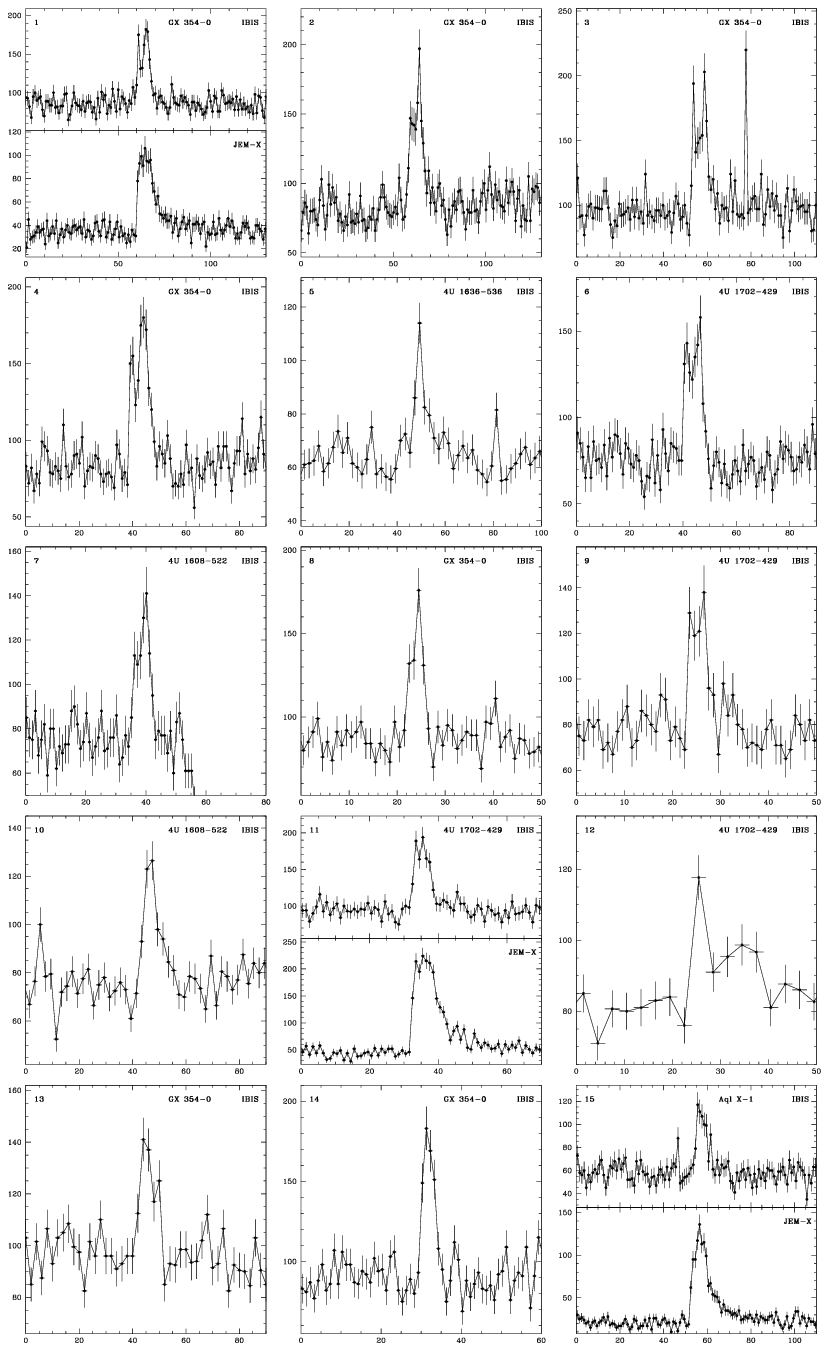}}
\end{figure}

\begin{figure}[hp]
\vspace{-5.5cm}
\centerline{\includegraphics[width=1.5\linewidth]{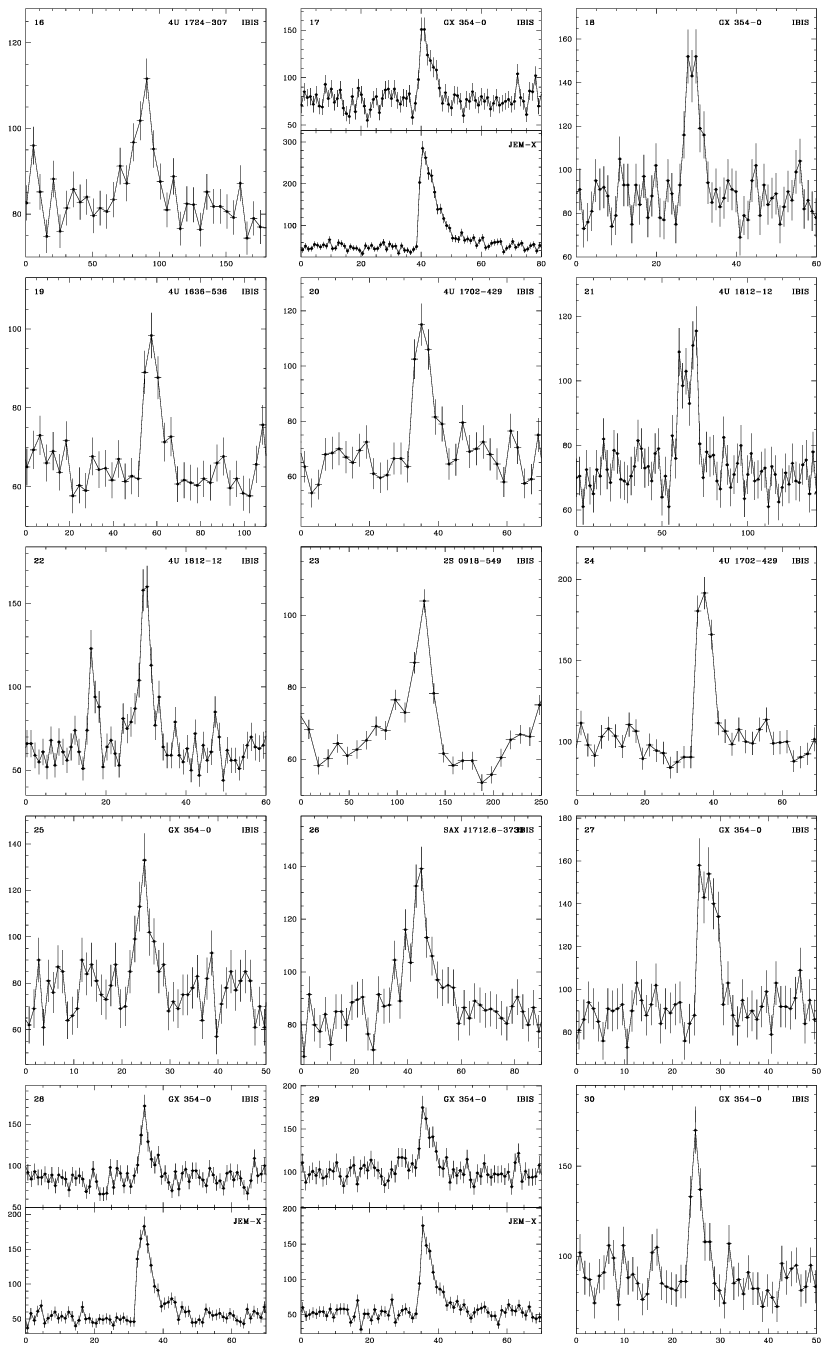}}
\end{figure}

\begin{figure}[hp]
\vspace{-5.5cm}
\centerline{\includegraphics[width=1.5\linewidth]{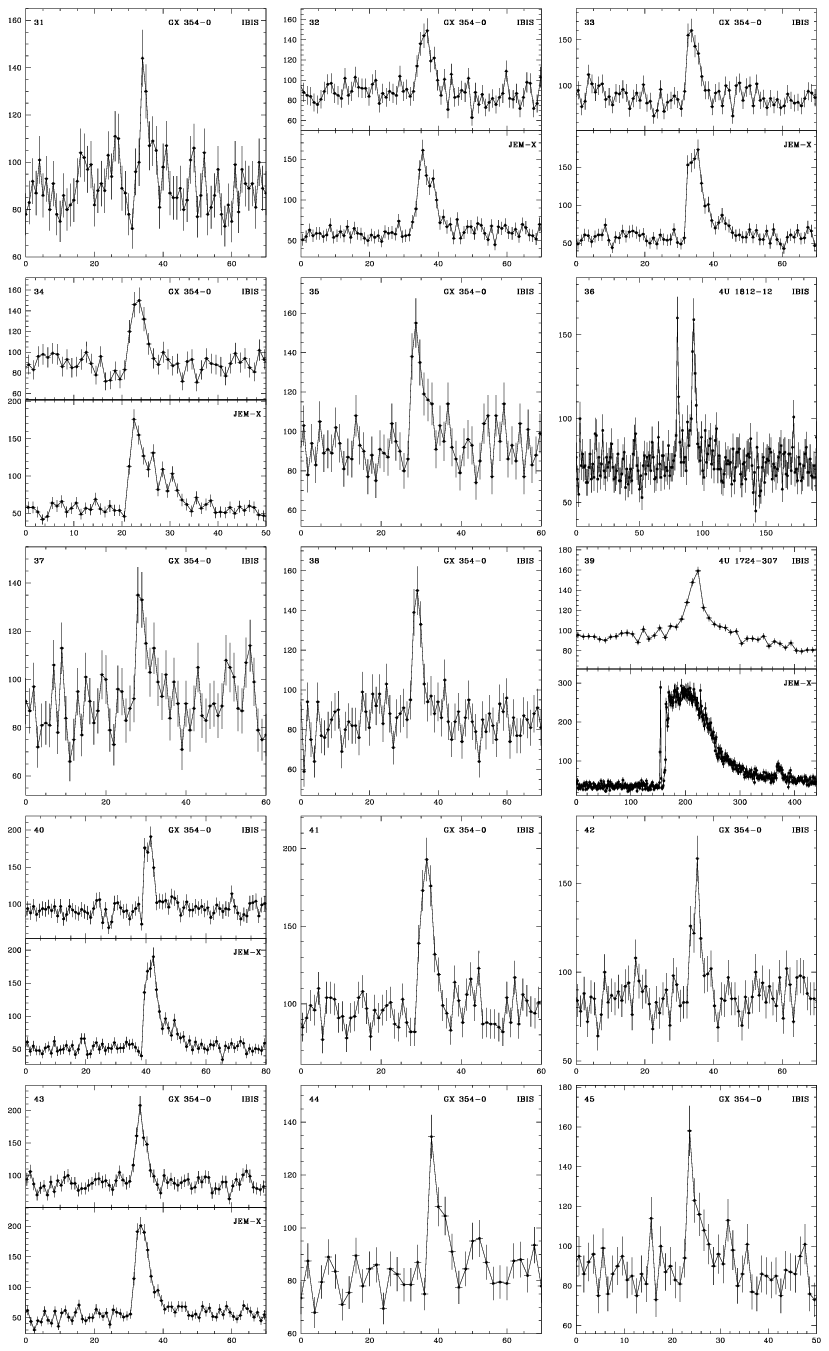}}
\end{figure}

\begin{figure}[hp]
\vspace{-5.5cm}
\centerline{\includegraphics[width=1.5\linewidth]{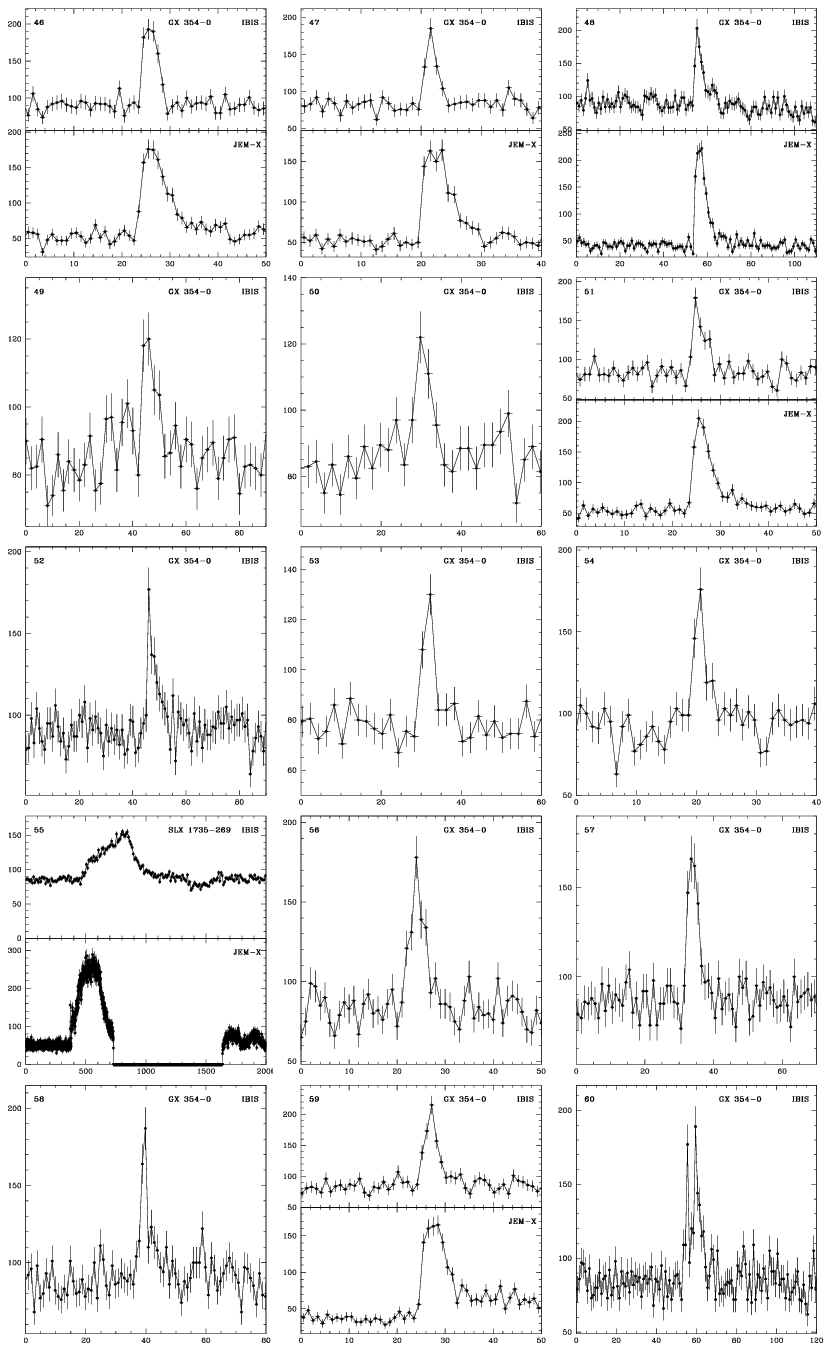}}
\end{figure}

\begin{figure}[hp]
\vspace{-5.5cm}
\centerline{\includegraphics[width=1.5\linewidth]{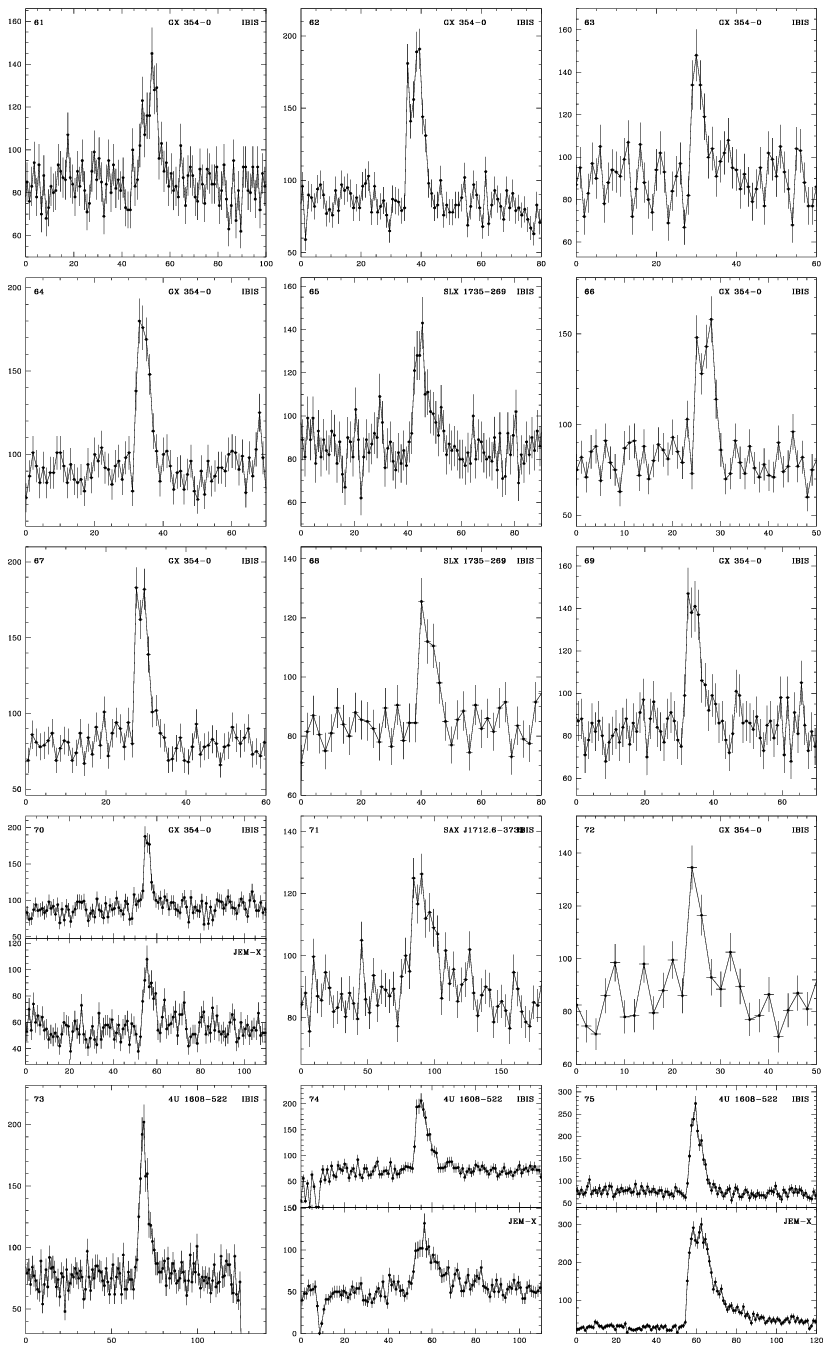}}
\end{figure}

\begin{figure}[hp]
\vspace{-5.5cm}
\centerline{\includegraphics[width=1.5\linewidth]{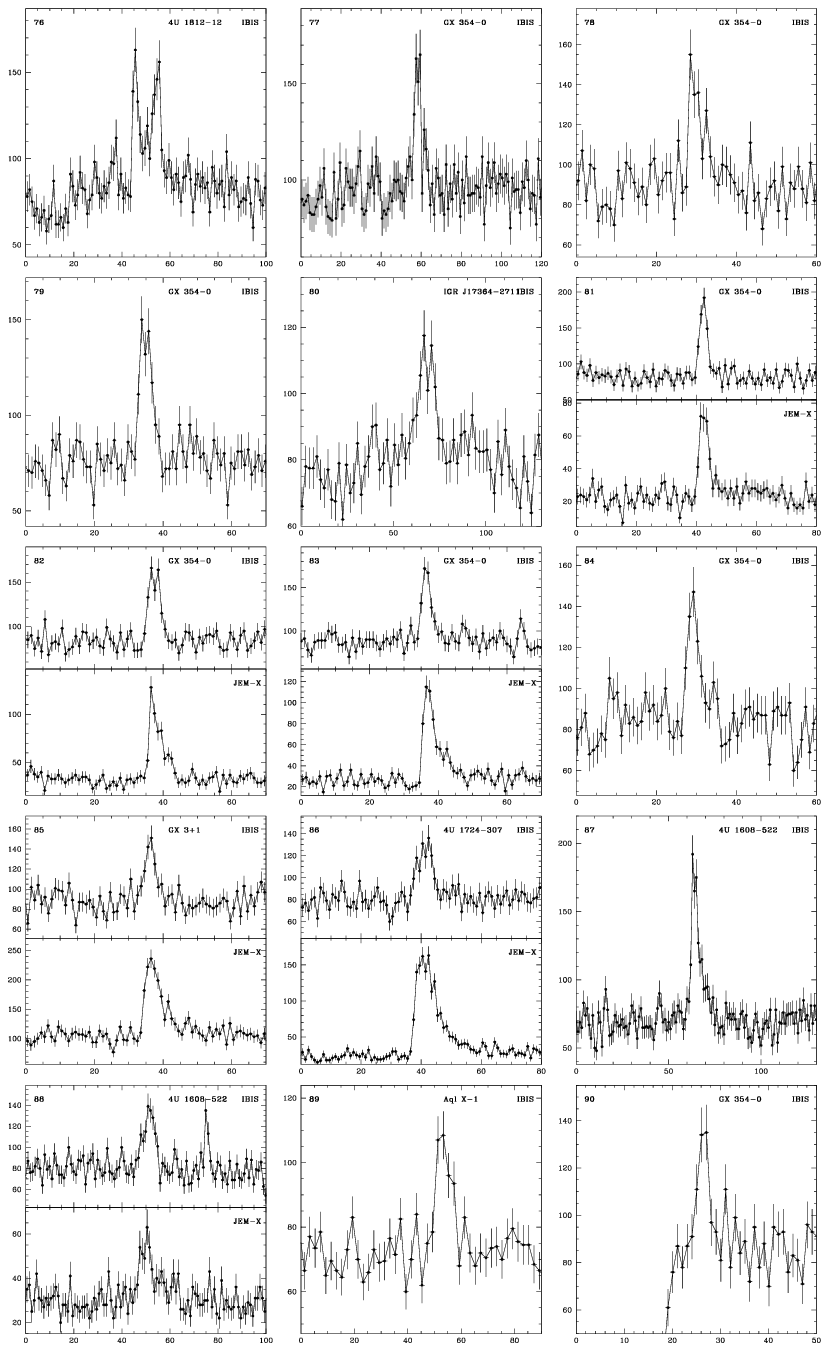}}
\end{figure}

\begin{figure}[hpt]
\vspace{-8.5cm}
\centerline{\includegraphics[width=1.5\linewidth]{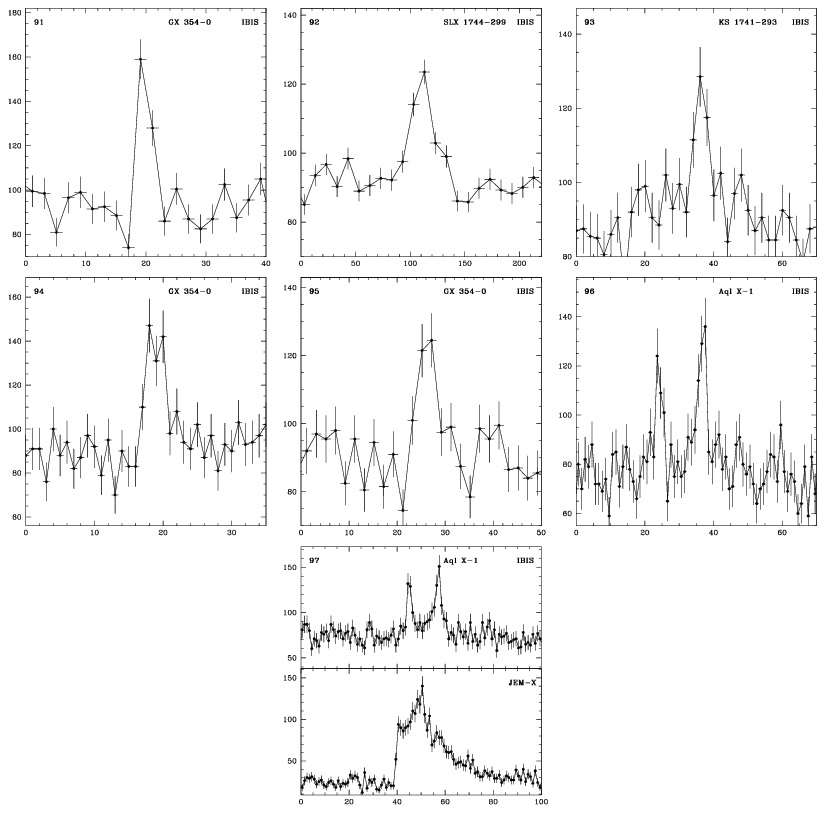}}
\vspace{-9.cm}
\caption{\rm Time profiles of 97 X-ray bursts detected by IBIS/ISGRI 
in the energy range 15--25 keV and identified with bursters.
For the cases where the burst was detected by the JEM-X monitor, 
the profile is also shown in the energy range 3--20 keV.}
\label{fig:burst_lc}
\end{figure}

\subsection*{A New X-ray Burster}
\noindent
The X-ray burst no. 80 in Table 2 was detected
by the IBIS/ISGRI telescope on February 17, 2004,
from a hitherto unknown burster located in the
Galactic center region (at $\sim27$\arcmin\ from the source
SLX 1735-269). Figure 4a shows an image of the
sky area in the field of view (a $S/N$ map) obtained in
the energy range 15-25 keV over 10 s during a burst.
The new source that we named IGR J17364-2711,
the only statistically significant source in this image,
was detected at $S/N=8.7$. Its position, $R.A.=17\uh36\um28\us$ 
and $Decl.=-27\deg11\arcmin56\arcsec$\ (epoch 2000.0),
was determined with an accuracy of 2 arcmin\footnote{
Because of a peculiarity of the IBIS coded mask, there is
duality in determining the position of the source in several
cases. The other possible position of the source of the
burst under consideration, $R.A.=17\uh38\um05\us$ and 
$Decl.=-37\deg49\arcmin05\arcsec$, is less likely than 
the above one, although it cannot be completely rejected.}. 
The maximum photon flux reached $1.6\pm0.3$ Crab, which
corresponded to a 15--25 keV luminosity $\simeq 8\times10^{37}$ 
\ergs\ at the distance of the Galactic center
$d=8.5$ kpc. The burst spectrum was very soft and
the photon flux at energies $\sim30$ keV fell below the
detection level. Assuming the spectrum to have a
Wien shape with a temperature $kT\simeq2.5-2.8$ keV
typical of bursts with photospheric expansion, we
find the bolometric luminosity of this burst, 
$L_{\rm B}\simeq (4-5)\times10^{38}$ \ergs. In complete 
agreement with the assumption of photospheric expansion, 
this luminosity is actually close to the Eddington limit. 
Unfortunately, the burst was not detected in the standard
X-ray energy range, since during the simultaneous
observations with the JEM-X monitor it was at
the very edge of the JEM-X field of view, which is
narrower than the IBIS one. Thus, it was not possible
to refine $kT$.

Figure 4b shows an image of the same sky area 
as that in Fig. 4a, but it was obtained over the entire
observing session of February 16--17, 2004, except
the pointing during which the burst occurred. In the
image, we clearly see four sources detected in this
region. There is no IGR J17364-2711 among them.
The corresponding $3\sigma$ limit on its persistent 
15--25 keV flux is 4 mCrab. Assuming the source to have a
power-law spectrum with an index $\gamma\simeq2.1$, we obtain
a fairly stringent limit on its 2--30 keV luminosity,
$L_{\rm X}\la1.2\times10^{36}$ \ergs. This source complements
the list of X-ray bursters (see, e.g., Cocchi et al. 2001;
Cornelisse et al. 2002) that have never been observed
in a state of persistent X-ray emission, but that have
been detected only during rather rare bursts.
Note that the nearest source detected in the soft
(0.1--2.4 keV) X-ray energy range during
the highly sensitive ROSAT all-sky survey,
1RXS J173602.0-272541, is located at an angular
distance of $\sim15$\arcmin\ from \mbox{IGR\,J17364-2711}.

\subsection*{Previously Known Bursters}
\noindent
In general, the bursts detected by the IBIS/ISGRI
telescope are harder and more intense than ordinary
events of this type. Therefore, it is quite pertinent to
provide here information about these sources.

{\bf 2S 0918-549} is a member of a very close X-ray
binary with a helium white dwarf as the donor. It was
identified as a burster by Jonker et al. (2001). Over the
entire history of its observations, seven type-I X-ray
bursts (in't Zand et al. 2005), including the very long
burst observed by the BeppoSAX satellite on October
1, 1996, for $\sim40$ min (the exponential decay time
was 117 s), have been detected from this source. The
remaining bursts did not exceed 6-29 s in duration.
The maximum fluxes for most of the bursts reached
2-3 (in the very long burst, 3.7) Crab in the standard
X-ray energy range. Two additional weak bursts from the source 
could not be classified as type-I bursts on the basis of the 
available data. Analysis of the IBIS/ISGRI
data revealed the tenth of the X-ray bursts ever detected
from this burster (Table2, burst 23). The maximum
15-25 keV flux was 3.6 Crab.

\begin{figure}[t]
\vspace{0.5cm}
\centerline{\includegraphics[width=1.0\linewidth]{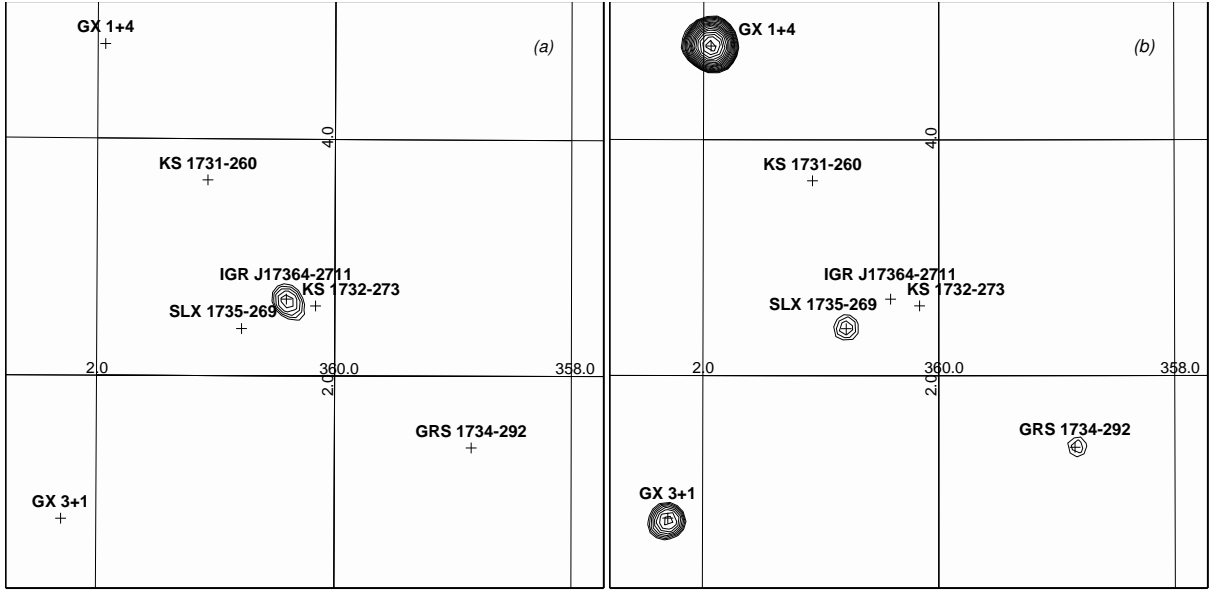}}
\caption{\rm Images of the sky area in the IBIS/ISGRI field of 
view obtained on February 17, 2004, (a) during burst no. 80 
detected from the hitherto unknown burster \mbox{IGR J17364-2711} and 
(b) during the entire observing session (except the burst time). 
The exposures were 10 s and more than 93000 s in the former and 
latter cases, respectively. Both images were obtained in the energy
range 15--25 keV. The contours indicate the region of reliable 
detection of the sources at the signal-to-noise ratio $S/N = 4,
4.7, 5.5, 6.5, 7.6, 8.9, . . .$ standard deviations (given with 
a logarithmic step of 1.17).}
\label{fig:igrimage}
\end{figure}

{\bf 4U 1608-522} is known as a soft X-ray transient
with a lifetime of 2--3 months and a recurrence time of
about one year. The first X-ray bursts fromtheNorma
constellation presumably associated with it were detected
by the Vela 5 and Uhuru observatories (Belian
et al. 1976; Grindley and Gursky 1976). Subsequently,
22 bursts were detected from 4U 1608-522
by the Hakucho observatory. Their analysis showed
a strong correlation between the burst properties and
the persistent flux from the source: the burst duration
in its on state was $\sim 8$ s, while in the off state it
varied between 10 and 30 s; the maximum burst flux
in the latter case was appreciably weaker (Murakami
et al. 1980a). Interestingly, although the burst rate
generally did not depend on the flux, the two bursts
detected in the off state occurred with an interval of
only $\sim 10$ min (Murakami et al. 1980b). Analysis of
the 17 bursts observed from 4U 1608-522 by the
Tenma observatory revealed the appearance of a hard
component in the source's spectrum both in the quiescent
state and during the burst, when the flux fell
below a certain level (Nakamura et al. 1989). The hard
component during the burst was explained by Comptonization
of the blackbody radiation in the high temperature
plasma surrounding the neutron star.
The double-peaked structure in the soft ($\le 6$ keV) energy
channels in the profile of the X-ray burst detected
from the source in 1987 by the EXOSAT observatory
was also explained by Comptonization. No photospheric
expansion with which this structure is usually
associated was detected in this burst (Gottwald et al.
1987). In 2005, a superburst with a duration of several
hours was detected from 4U 1608-522 (Remillard
and Morgan 2005; Kuulkers 2005).

Table2 lists the seven bursts detected by the IBIS
telescope from this source: two in March 2003, two
in March 2004, and three in September 2003 (bursts
7, 10, 73, 74, 75, 87, and 88). Their detection in
the ISGRI energy range is indicative of a fairly large
hardness, suggesting that they occurred during the
source's low state. The short burst duration of 8-14 s, 
which is more characteristic of bursts in the
source's high state, can again be associated with the
hard energy range in which the observations were
performed. A decrease in duration in the hard energy
channels compared to the soft ones is typical of type-I 
bursts. The bursts had a classical shape: a fast rise
and a slower exponential decay. In the standard X-ray
energy range (three events, 74, 75, and 88, were also
detected by the JEM-X monitor), a double-peaked
structure appeared in the burst profiles near the flux
maximum.

{\bf 4U 1636-536} has been known as an X-ray
burster since 1976, when several bursts were detected
from it by the OSO-8 satellite (Swank et al.
1976a). Owing to their high rate and regularity, the
bursts from this source were analyzed by all of the
succeeding X-ray observatories, including SAS-3,
Hakucho, Tenma, EXOSAT, KVANT, RXTE, and
BeppoSAX. A significant variability (see, e.g., Lewin
et al. 1987) both in the burst profile shape and peak
flux (by a factor of 6) and in the recurrence time (by
a factor of 24) was found. The IBIS/ISGRI telescope
detected two bursts of moderate intensity $\sim 1$ Crab
and duration $\sim 10$ s (bursts 5 and 19 in Table 2) from
this source.

{\bf 4U 1702-429} was identified as a burster in 1979
based on observations from the Hakucho satellite
(Makishima et al. 1982). Bursts from it may have also
been observed previously by the OSO-8 (Swank et al.
1976b) and SAS-3 (Marshall et al. 1977) satellites.
Subsequently, bursts from the source were analyzed
by the EXOSAT (Oosterbroek et al. 1991), KVANT
(Emel'yanov et al. 2001), GRANAT(Grebenev et al.
2000), RXTE (Markwardt et al. 1999), and BeppoSAX
(Cornelisse et al. 2003) observatories. Analysis
of the source's burst activity revealed its variations
on a time scale of about one year.
Six bursts detected by the IBIS/ISGRI telescope
(Table 2, bursts 6, 9, 11, 12, 20, and 24) were identified
with 4U 1702-429. The first three bursts had a
double-peaked feature in the profile near the flux maximum
(Fig. 3). For burst 11 observed by the JEM-X
monitor, such a structure was also found in the profile
in the standard X-ray energy range. These profiles
were generally morphologically similar and the burst
began almost simultaneously in both energy ranges.
On the other hand, the peak flux was reached earlier
and the burst lasted longer in the standard energy
range.

{\bf The X-ray transient SAX~J1712.6-3739} was
discovered by the BeppoSAX observatory on August
24, 1999, at the time of its outburst (in't Zand
et al. 1999). The weak ($\sim 1.6$ mCrab in the energy
range 0.1--2.4 keV) X-ray source 1RXS~J171237.1-373834 
from the ROSAT all-sky survey fell within the
error circle. Several days later, on September 2, 1999,
a type-I X-ray burst was detected by BeppoSAX
from this direction after the fading of the transient
(Cocchi et al. 1999). This burst has long remained the
only known one. Using the IBIS/ISGRI telescope,
we were able to detect two more bursts from the
source (nos. 26 and 71 in Table 2). Both bursts had
a triangular profile. We failed to trace the evolution of
the bursts in the standard X-ray energy range, since
the source was outside the JEM-X field of view in
both cases.

{\bf The burster 4U 1724-307} is located in the globular
cluster Terzan~2. Grindley (1978) showed that
4U~1724-307 was the source of the intense and long
($\ga300$ s) burst detected in 1975 by the OSO-8 satellite
(Swank et al. 1977). At least three additional intense
bursts similar to the 1975 burst with strong photospheric
expansion have been detected to date from
the source (by the ART-P telescope onboard the
GRANAT observatory, the PCA instrument onboard
the RXTE observatory, and the MECS instrument
onboard the BeppoSAX observatory; Guainazzi et al.
1998; Molkov et al. 2000). More than twenty weaker
bursts have also been detected.

In Table 2, three bursts (16, 39, and 86) are associated
with this source. Burst 39 supplements the
list of its known superstrong bursts. Since the source
was at the center of the IBIS and JEM-X field of
view at the time of this burst, Fig. 3 shows the time
profiles of this event in two energy ranges. We see
from this figure that they differ greatly: whereas in the
soft energy range the burst had a narrow precursor
and a broad main event with a sharp rise and a long
($\sim 200$ s) exponential decay, i.e., had all of the features
characteristic of type-I bursts (Hoffman et al. 1977),
in the hard energy range it had an almost triangular
profile, the flux reached its maximum considerably
later than in the soft energy range, and the burst itself
lasted only $\sim45$ s. Note that burst 86, which was also
detected by JEM-X, in the soft energy range has a
profile that does not differ greatly from its IBIS profile.

{\bf GX~354-0 (4U~1728-337)} clearly stands out
among the bursters from which the IBIS/ISGRI telescope
detected X-ray bursts. This source is responsible
for 61 bursts in Table 2. Discovered in 1976 (Lewin
1976; Hoffman et al. 1976), it rapidly became one of
the best known bursters. Its burst recurrence time
ranges from several hours to several tens of hours.
The large number of bursts detected from \mbox{GX 354-0}
by the IBIS/ISGRI telescope allowed the statistical
distributions of its bursts in recurrence time, duration,
and maximum flux to be analyzed.

\begin{figure}[t]
\centerline{\includegraphics[width=0.9\linewidth]{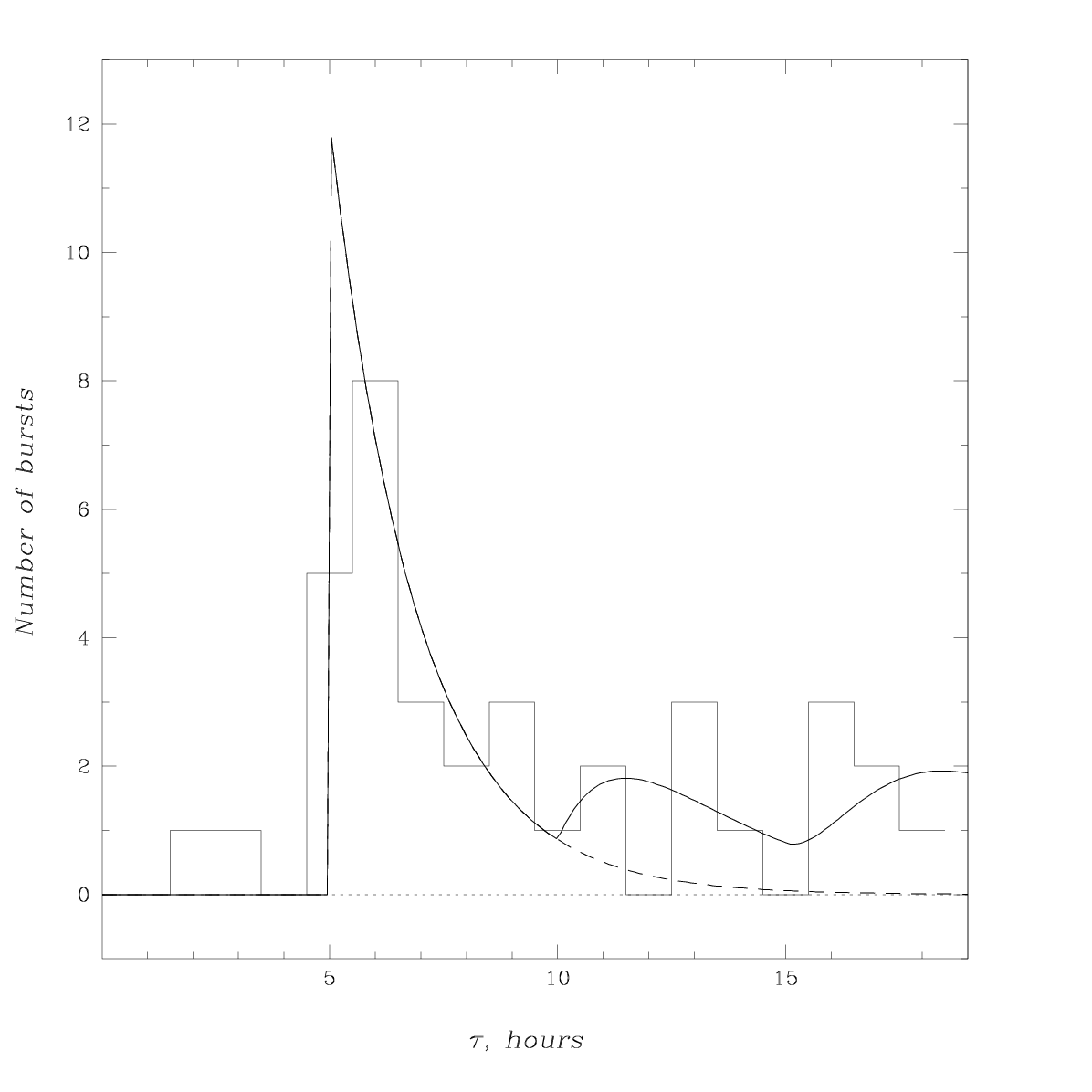}}
\caption{\rm Distribution of bursts from the X-ray burster
\mbox{GX 354-0} in recurrence time and its model fit described
in detail in the text.}
\label{fig:GX354_number-recur}
\end{figure}

Figure 5 shows the distribution of X-ray bursts
in recurrence time $\tau$ (the interval between the detection
times of two successive events). We took into
account only 36 events with $\tau\leq20$~h. Since the
time of continuous observations of the source usually
did not exceed 20~h, bursts that were not detected
by the detector are almost certain to have occurred
between more widely spaced events. Assuming that
the properties of the continuum emission from the
source (e.g., the accretion rate) did not change and
that the distribution of bursts in $\tau$ generally obeyed
Poissonian statistics, we fitted it by the formula

\[
\frac{dN}{d\tau} = N_0 \lambda
e^{-\lambda(\tau-\tau_0)}\theta(\tau-\tau_0).
\]
Here, the function
\[ 
\theta(x)=\left\{
\begin{array}{ll}0
, &\mbox{if } x<0 \\
1, &\mbox{if } x\geq 0 
\end{array}
\right.
\]

takes into account the threshold nature of thermonuclear
flashes (the next burst can occur only after the
time $\tau_0$ it takes to accrete a critical mass of matter
on the stellar surface) and $\lambda$ specifies the mean burst
rate. We additionally took into account the fact that
some of the bursts could be missed. Therefore, the
events corresponding to double

$$
\frac{dN^{(2)}}{d\tau} = a_2 N_0 \lambda^2 (\tau-2\tau_0)
e^{-\lambda(\tau-2\tau_0)}\theta(\tau-2\tau_0)
$$
and triple the recurrence time
$$
\frac{dN^{(3)}}{d\tau} = a_3 N_0 \lambda^3 \frac{(\tau-3\tau_0)^2}{2}
e^{-\lambda(\tau-3\tau_0)}\theta(\tau-3\tau_0).
$$ 
may be present in Fig. 5.

As the solid curve in Fig. 5 shows, the above
formulas provide satisfactory agreement with the observational
data. The $\chi^2$ value normalized to $N=15$
degrees of freedom is $\chi^2_N=1.17$. The best-fit parameters
are $\lambda=0.53\pm0.05$ h$^{-1}$ and $\tau_0=5.0\pm0.7$ hours. The
fraction of the events that were mistaken for successive
ones proved to be fairly large, $a_2=0.3\pm0.1$ and
$a_3=0.5\pm0.1$ (their contribution is indicated by the
dashed line in the figure). The mean burst recurrence
time for this source is $<\!\tau\!>=\tau_0+1/\lambda\simeq6.9$ hours.

\begin{figure}[t]
\centerline{\includegraphics[width=0.8\linewidth]{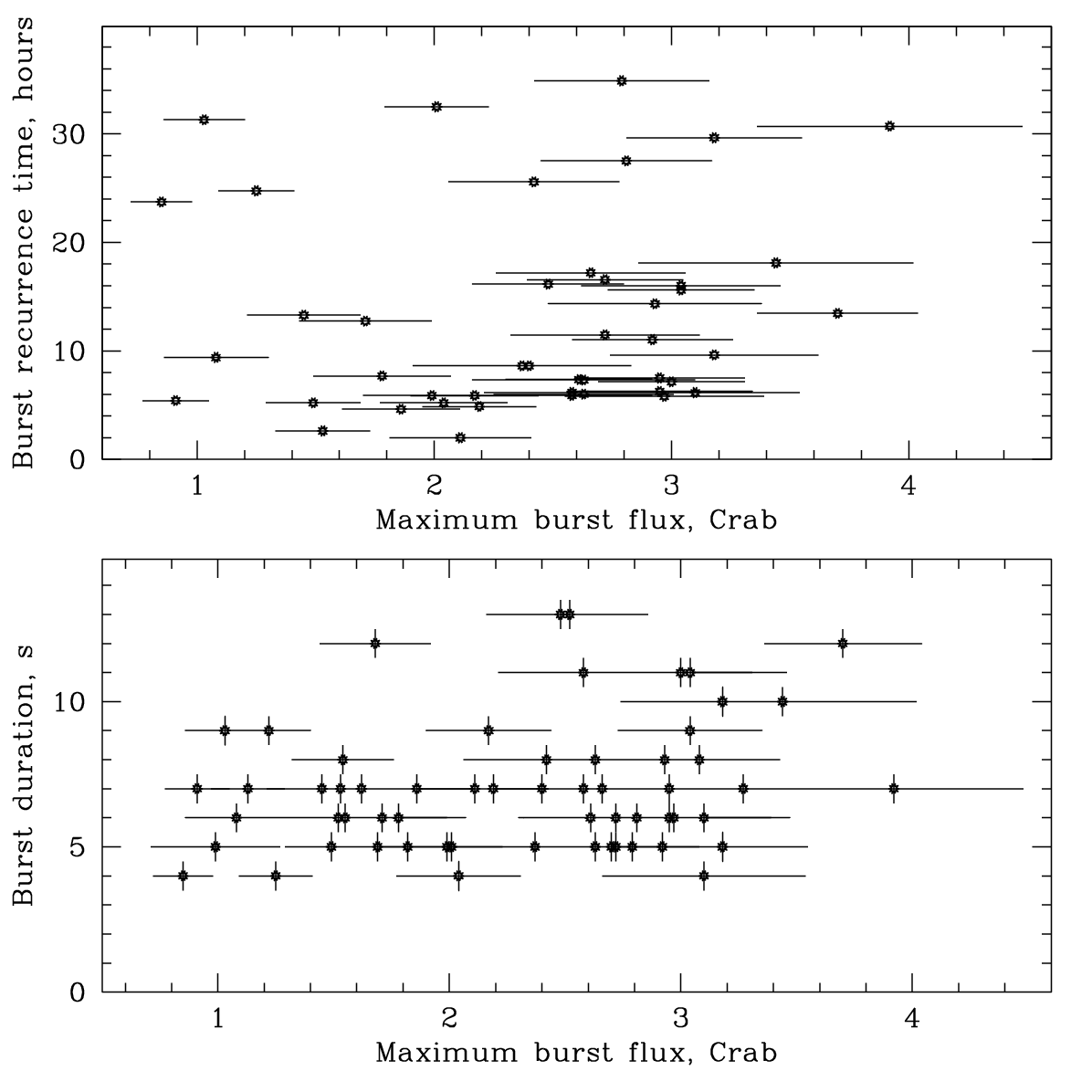}}
\caption{\rm Recurrence time and duration of the X-ray bursts 
detected from the burster \mbox{GX 354-0} vs. maximum flux during the
burst.}
\label{fig:GX354_dur_recur-bmax}
\end{figure}

In Fig. 6, burst recurrence time $\tau$ (upper panel)
and duration $T$ (lower panel) are plotted against maximum
flux $F$. We failed to find any regularities in
the relationship between these quantities. However,
if we examine the relationship between the burst fluence
(a quantity equal to the product of the mean
flux and the duration, $E= <\!F\!>T$) and the recurrence
time $\tau$, a certain regularity shows up (Fig. 7). We
fitted it by a linear law $E=E_0+A\tau$. This behavior
seems natural: the longer the interval between the
bursts is, the more matter is accumulated on the neutron
star surface and the more intense the coming burst is. It may
even be assumed that $A\simeq <\!F_X\!>/\alpha,$ where 
$<\!F_X\!>$ is the mean continuum flux from the source 
(in units of the flux from the Crab Nebula) and $\alpha\sim 100$ 
is a factor that shows the extent to which the energy
release via accretion is more efficient than the energy
release through thermonuclear burning. The situation
was slightly complicated by the presence of events
corresponding to multiple recurrence times in the
sample. To reduce their influence, we included only
events with $3\ \mbox{\rm h}\leq\tau\leq10$ h in our fit. 
The derived parameters are $E_0=2.36\pm1.57$~Crab~s and 
$A=0.67\pm0.23$ Crab s h$^{-1}$. The corresponding mean flux
is $<\!F_X\!>\simeq A\alpha\simeq 19$ mCrab.

\begin{figure}[t]
\centerline{\includegraphics[width=0.8\linewidth]{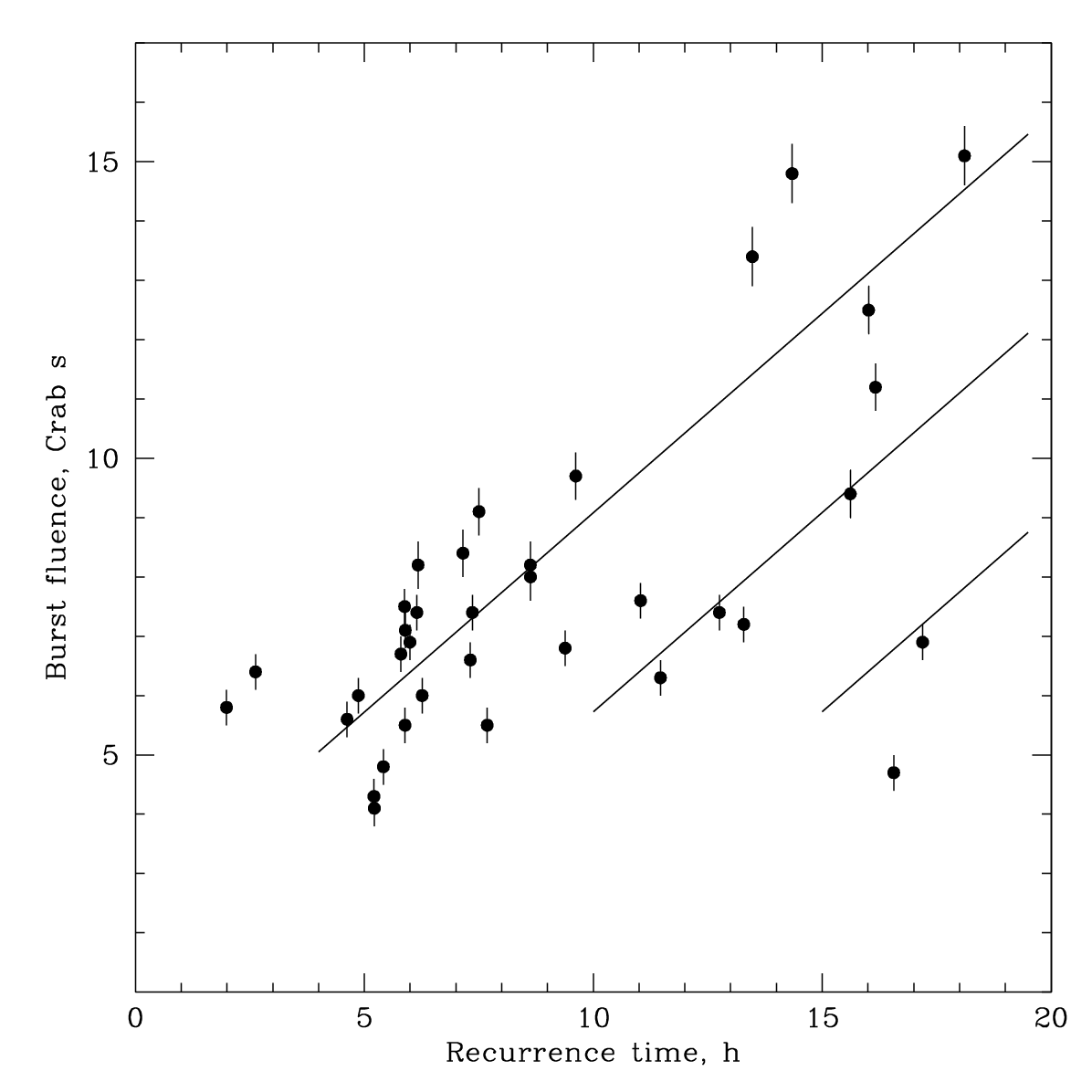}}
\caption{\rm Fluence of the bursts detected from the burster
\mbox{GX 354-0} vs. duration of the quiescent period preceding
the burst and its linear fit. Some of the data points corresponding
to long recurrence times and low fluences are associated with the 
misses of the bursts occurred during the INTEGRAL slewing.}
\label{fig:GX354_fluence-recur}
\end{figure}

In Fig. 8, burst fluence is plotted against burst
duration $T$. Here, we also see a regularity whose
fitting by a linear law $E=E_1+B\, T$ yielded 
$E_1=2.61\pm0.60$ Crab s and $B=0.61\pm0.08$ Crab s s$^{-1}$.
Such a linear dependence naturally arises in the case
of intense bursts with near-Eddington fluxes, but
it apparently remains qualitatively valid for weaker
bursts as well.

\begin{figure}[t]
\centerline{\includegraphics[width=0.8\linewidth]{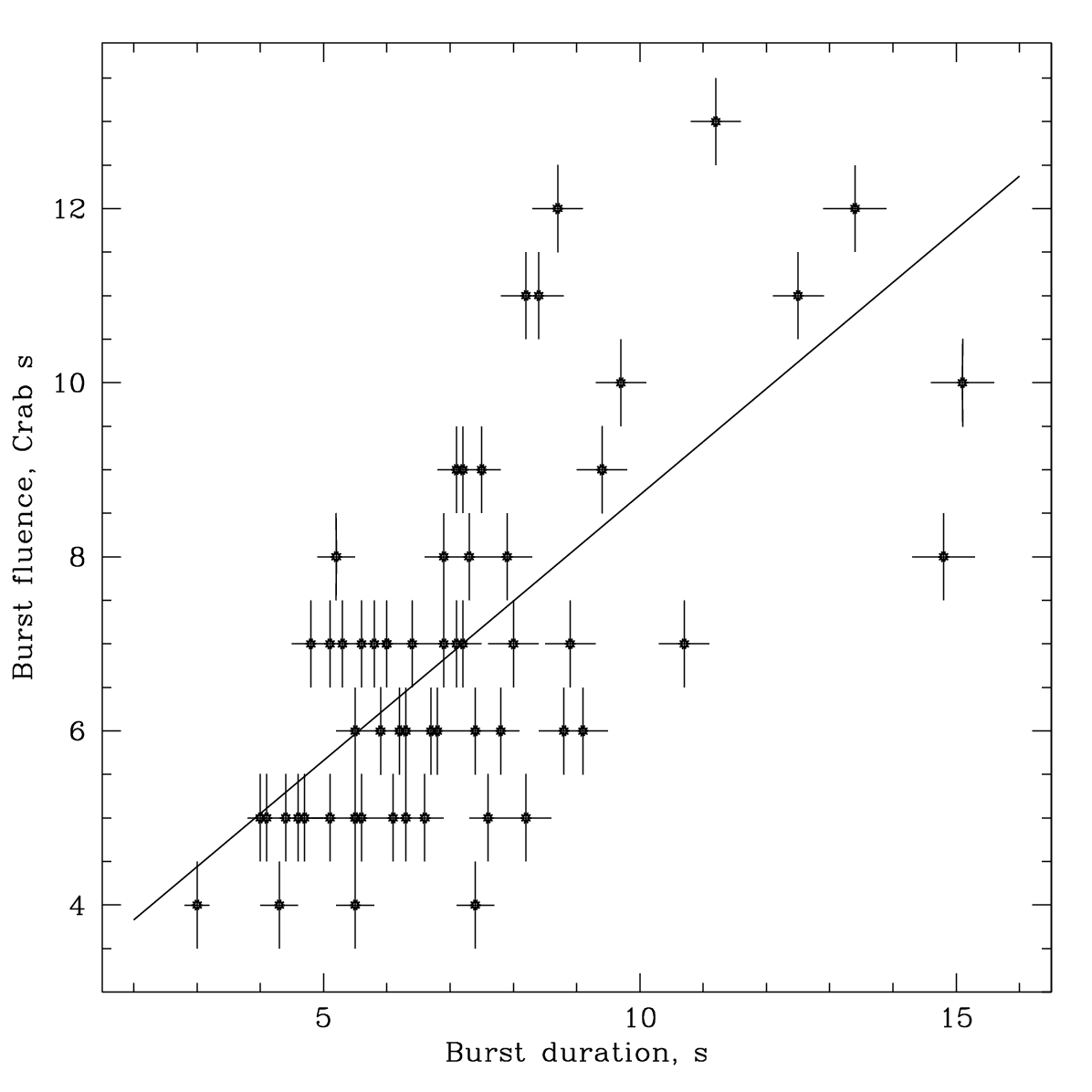}}
\caption{\rm Fluence of the bursts detected from the burster \mbox{GX 354-0}
vs. burst duration and its linear fit.}
\label{fig:GX354_fluence-dur}
\end{figure}

Figure 9 shows the distributions of the detected
bursts in maximum flux $F$ and duration $T$. On the
whole, the bursts are distributed in flux fairly uniformly
up to 3.2 Crab. An excess of bursts is observed
in the regions of maximum fluxes $\sim1.5$ and
2.2--3.2 Crab. At this juncture, it is probably premature
to talk about the existence of two types of
bursts characteristic of this source. The distribution
of bursts in duration is more nonuniform. Most of the
bursts have durations in the range 5--7 s. Using the
above dependences $dN/d\tau$, $E(\tau)$, and $E(T)$, we can
attempt to approximate the distribution of the detected bursts in
duration:

$$
\frac{dN}{dT}=\frac{dN}{dE} B= \frac{dN}{d\tau}\ \frac{B}{A}, 
$$
where $\tau=(B\,T+E_1-E_0)/A.$. The result of this approximation is 
indicated in Fig. 9 by the solid line.We had to correct the normalization
of this distribution to allow for the difference in
the numbers of bursts presented in Figs. 5 and 9.

\begin{figure}[t]
\centerline{\includegraphics[width=0.8\linewidth]{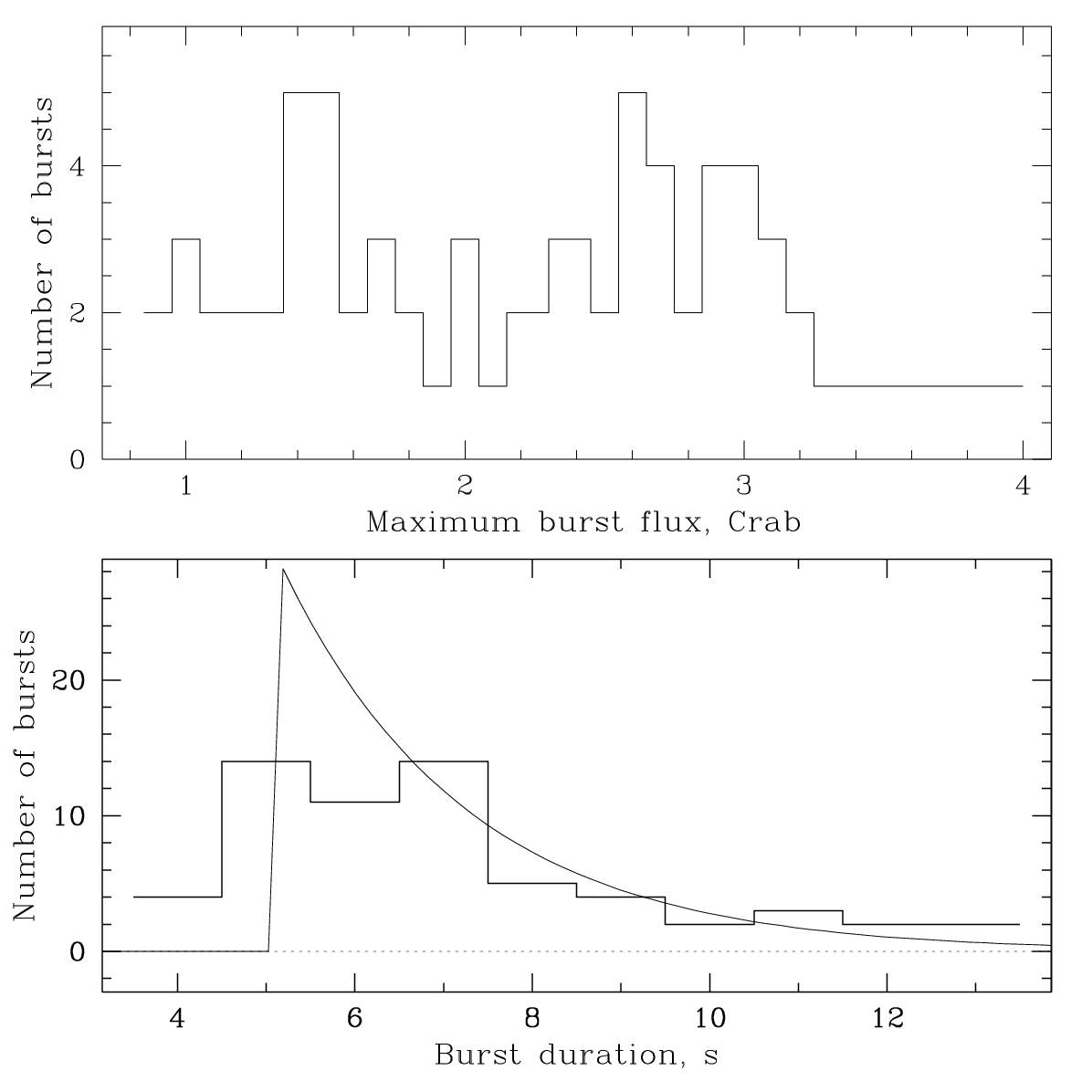}}
\caption{\rm Distributions of bursts from the X-ray burster
\mbox{GX 354-0} in (top) burst maximum flux and (bottom) burst duration.
In the latter case, the solid curve also indicates the
model fit to the observed distribution (see the text).}
\label{fig:GX354_number-bmax_dur}
\end{figure}

{\bf SLX 1735-269} has a hard power-law X-ray spectrum.
As a result, the nature of its compact object has
long remained uncertain (see, e.g., Grebenev et al.
1997). The first X-ray burst from this source that put
an end to the debate about its nature was detected
only recently (Bazzano et al. 1997). In Table 2, this
burster is responsible for bursts 55, 65, and 68. The
properties of the first burst with an unusually long
duration ($\sim 450$ s in the IBIS/ISGRI energy range
and more than $1500$ s in the JEM-X energy range)
were studied in detail by Molkov et al. (2005). The
burst was so intense that it led to a saturation of the
JEM-X telemetry channel and the filtering switchon.
This is why its profile has an unusual shape in
the standard X-ray energy range in Fig. 3. Molkov
et al. (2005) provided a reconstructed X-ray profile of
this burst. Bursts 65 and 68 lasted $\sim 10$ s and were
considerably less intense than the first burst.

{\bf The transient KS~1741-293} was discovered and
identified as a burster (two bursts) by the KVANT
module during its outburst of August 20--22, 1989
(in't Zand et al. 1991). It was at the very edge of the
error circle for the source of the X-ray bursts detected
by the SAS-3 satellite from the Galactic center region
(MXB~1743-29; Lewin et al. 1976). This may
be the same source. However, since the position of
MXB~1743-29 is inaccurate and the region is densely
populated, a chance coincidence seems more likely.
Besides, as in't Zand et al. (1991) pointed out, the
time profiles of their bursts and the bursts detected
by the SAS-3 satellite differ greatly.

KS~1741-293 was in a higher state during the
INTEGAL observations of the Galactic center region
in the spring of 2003 (Belanger et al. 2004) and the
spring of 2004 (Grebenev et al. 2004). The only burst
identified with it in Table 2 (burst 93) was detected on
March 30, 2004.

{\bf GX 3+1 (4U 1744-265)} is one of the brightest
X-ray sources in the field of the Galactic center. It has
been known as a burster since 1983 (Makishima et al.
1983). The IBIS/ISGRI telescope detected one burst
(no. 85 in Table 2) of moderate intensity from it with
an almost triangular profile. In the soft X-ray energy
range (the burst was also detected by the JEM-X
monitor), its profile was closer to the canonical profile
of a type-I burst (Fig. 3).

{\bf SLX 1744-299} was discovered by the Spacelab-2 
observatory in 1987 (Skinner et al. 1987). Subsequently,
an X-ray burst was detected from this
direction (Skinner et al. 1990); it showed that, in
fact, a pair of close (separated by only $\sim2.8$\arcmin) sources,
SLX 1744-299 and SLX 1744-300 (the latter is a
burster), is hidden here. In 1990, five more bursts,
including a very intense and long ($\sim300$~s) burst, were
detected during intensive observations of the Galactic
center region by the ART-P telescope onboard the
GRANAT observatory (Pavlisnky et al. 1994). The
long burst was identified with SLX 1744-299; i.e.,
the first source also turned out to be a burster. The
IBIS/ISGRI telescope detected burst 92 in Table 2
from these sources.

{\bf 4U 1812-12} was first identified as a burster by
Murakami et al. (1983). Since then and until the
INTEGRAL observations, a total of 12 bursts were
detected from it (Fleischman et al. 1985; Cocchi et al.
2000). Four of the bursts detected by IBIS/ISGRI
were associated with this burster (bursts 21, 22, 36,
and 76 in Table 2). All these bursts had a distinct
double-peaked profile structure; bursts 22, 36, and
76 may be said to have two separate peaks of equal
amplitude spaced 15-20 s apart. Similar profiles were
observed for most of the bursts detected in 1996-1999 
by the BeppoSAX observatory (Cocchi et al.
2000). This was explained by the attainment of a
super-Eddington luminosity during the burst and
was used to estimate the distance to the source,
$\simeq 4$ kpc.

{\bf The transient burster Aql X-1 (4U 1908+005)} 
has much in common with 4U 1608-522
discussed above. X-ray bursts from it were first
observed in May 1980 by the Hakucho observatory
(Koyama et al. 1981). The IBIS/ISGRI telescope
detected four bursts from this burster during the
period of observations under consideration (bursts
15, 89, 96, and 97 in Table 2). Bursts 15 and 97
were also observed by the JEM-X monitor. The
detection of burst 15 by JEM-X was first reported
by Molkov et al. (2003), who discussed in detail the
INTEGRAL observations of the X-ray outburst of the
source occurred in the spring of 2003. These authors
also mentioned the detection of another X-ray burst
$\sim 2.5$ h after the first burst. Our analysis shows that
this burst was too soft to be detectable by ISGRI.
The time profiles of bursts 96 and 97 in the hard
energy range have a double-peaked structure with a
detached precursor. In the soft energy range (at least
for burst 97), the precursor merges with the main
event and the maximum flux is reached in the interval
between the two hard peaks. The burst duration in the
X-ray energy range appreciably exceeded its duration
at hard energies.

\subsection*{Other Burst Sources}
\noindent
We were able to associate 18 of the localized bursts
with sources unrelated to bursters. Given the peculiarities
of this work, we only mention them. One, two,
and five bursts were detected from the microquasar
\mbox{GRS 1915+105}, the pulsar \mbox{Vela X-1}, and the source
\mbox{4U 1700-377}, respectively. Four bursts were identified
with the soft gamma repeater \mbox{SGR 1806-20}
(see also Gotz et al. 2005). Six bursts were identified
with GRBs occurred in the IBIS field of view (they
were previously detected by the IBAS system). A total
of 10 GRBs were detected in the IBIS field of view
over the period of our observations, but the statistical
significance of detecting the four remaining bursts in
the light curves in the fairly soft energy range under
consideration was lower than $s_0$. As we mentioned
above, 69 more events detected by ISGRI were also
detected by the anticoincidence shield of the SPI
gamma-ray spectrometer, but were not detected by
the GOES monitors, i.e., they are probably GRBs
that occurred outside the IBIS field of view.

Even after the inclusion of all these events and
145 probable solar flares (events recorded by the
GOES monitors), 748 bursts that were detected by
IBIS/ISGRI, but were not unidentified remain. Some
of them are undoubtedly associated with instrumental
effects, for example, with the excitation of ISGRI
elements for a short time. Many bursts turned out to
be too weak to be reliably revealed in the image. We
plan to continue our analysis of unidentified bursts
and to present its results in a separate paper.

\section*{DISCUSSION}
\noindent

The main goal of this study was to search for
new X-ray bursters in a low (off) state and transient
bursters with low persistent luminosities. We
discovered one hitherto unknown burster named
\mbox{IGR J17364-2711} during its burst.

Is this number large or small for a $\sim 3.4$ Ms exposure
of the Galactic center region (see Fig. 1)? First,
let us estimate the burst recurrence time for the new
source:

$$
 \tau= \alpha T F/F_X \simeq 6\times 10^8\ \mbox{\rm s}\simeq 20\
 \mbox{\rm years}. 
$$ Here, $\alpha\sim100$ is a factor that shows the efficiency
of energy release via accretion compared to energy release via
thermonuclear burning on the neutron star surface. We assumed
all of the bursts from this source to be similar to the burst
detected by the \mbox{ISGRI} detector, i.e., to have a duration
$T\simeq 13$ s and a maximum flux $F\simeq 1.6$ Crab, and took
the persistent photon flux from the source to be $F_X\simeq
3\times10^{-6}$ Crab in the energy range 15--25 keV (which
corresponds to a typical (for low-mass X-ray binaries) off-state
2--30 keV luminosity $L_X\simeq1\times10^{33}$ \ergs\ at a
distance of 8.5 kpc for a spectrum similar to that of the Crab
Nebula). The detection of only one burst in $\sim 3.4$ Ms of
observations implies that there can be no more than 180 such
binaries in this Galactic region where we simultaneously observe
up to 80\% ($4\times10^9\ M_{\sun}$ in a
$\sim8^{\circ}\times8^{\circ}$ field) of the Galactic bulge
stars, not counting the stars ($3\times10^9\ M_{\sun}$) of other
Galactic components (see Grebenev et al. 1996). This number is
comparable to the number of persistent and transient sources,
including the bursters, known in the field. More severe
constraints on the number of unknown bursters that are
constantly in a low (off) state can be obtained using the X-ray
luminosity function of the Galaxy (see, e.g., Emel'yanov et al.
2001). However, given the hard energy range in which bursts were
observed and the related selection of their sources, caution
should be exercised here.

Indeed, the X-ray bursts detected by the IBIS/ISGRI
telescope and presented in Table 2 differ not so much
by the high intensity as by the spectral hardness,
which allows their sources to be separated into a
special group of bursters. During most of the observations,
other known bursters with a high and stable
burst rate (\mbox{GS1826-238}, \mbox{A1742-294},
\mbox{4U1820-303}, \mbox{4U1705-440}, \mbox{4U1735-444}, and others) were
constantly within the IBIS field of view. No burst
was detected from them. At the same time, for
example, 61 bursts were detected from \mbox{GX 354-0},
which accounts for more than half of all the ISGRI
events identified with bursters. This result cannot be
explained by the selection effect; it suggests that the
behavior of the source itself is peculiar.

Many of the detected bursts have a double-peaked
time profile (see Fig. 3). Generally, this suggests that
a critical Eddington flux is reached during a thermonuclear
explosion on the neutron star surface; as
a result, the atmosphere of the neutron star begins
to outflow and its photosphere begins to expand. The
expansion of the photosphere is accompanied by its
cooling. This gives rise to a dip in the time profile,
which is particularly large at hard energies, and to
a double-peaked structure. This picture is complicated
by Comptonization effects, which increase in
importance in a radiation-dominated medium, and by
peculiarities of the thermonuclear reactions. Unfortunately,
in most cases, we failed to obtain the burst
time profiles in the standard X-ray energy range. The
JEM-X field of view is only $\sim 15$\% of the IBIS field
of view. Therefore, the JEM-X monitor was able to
observe only 28 of the 97 IBIS bursts under consideration.

\section*{ACKNOWLEDGMENTS}
\noindent
We wish to thank S.V. Molkov and R.A. Krivonos
for helpful discussions. This work is based on the
observational data obtained by the INTEGRAL
observatory and provided through the Russian and
European INTEGRAL Science Data Centers. The
study was supported by the Russian Foundation
for Basic Research (project no. 05-02-17454), the
Presidium of the Russian Academy of Sciences (the
Nonstationary Astronomical Phenomena Program),
and the Program of the Russian President for
Support of Leading Scientific Schools (project no.
NSh-2083.2003.2).

\end{document}